\documentclass[aps,prd,twocolumn,groupedaddress,floatfix]{revtex4}

\usepackage[dvips]{graphicx}
\usepackage{amsmath,amssymb}
\usepackage{hyperref}
\usepackage{bm}
\usepackage{longtable}

\newcommand{\norm}[1]{\left| #1 \right|}

\newcommand{\upto}{\ \mbox{-} \ }

\newcommand{\der}[2]{\frac{\partial #1}{\partial #2}}
\newcommand{\uvec}[1]{\bm{\hat{#1}}}
\newcommand{\dvec}[1]{\dot{\bm{#1}}}
\newcommand{\duvec}[1]{\dot{\bm{\hat{#1}}}}

\begin{document}


\title{Parameter estimation for coalescing massive binary black holes with LISA
using the full 2-post-Newtonian gravitational waveform and spin-orbit
precession}


\author{Antoine Klein}
\email{aklein@physik.uzh.ch}
\affiliation{Institut f\"ur Theoretische Physik, Universit\"at Z\"urich,
Winterthurerstrasse 190, 8057 Z\"urich}

\author{Philippe Jetzer}
\affiliation{Institut f\"ur Theoretische Physik, Universit\"at Z\"urich,
Winterthurerstrasse 190, 8057 Z\"urich}

\author{Mauro Sereno}
\affiliation{Institut f\"ur Theoretische Physik, Universit\"at Z\"urich,
Winterthurerstrasse 190, 8057 Z\"urich}


\date{\today}

\begin{abstract}
Gravitational waves emitted by binary systems in the inspiral phase carry a
complicated structure, consisting in a superposition of different harmonics of
the orbital frequency, the amplitude of each of them taking the form of a
post-Newtonian series. In addition to that, spinning binaries experience
spin-orbit and spin-spin couplings which induce a precession of the orbital
angular momentum and of the individual spins. With one exception, previous
analyses of the measurement accuracy of gravitational
wave experiments for comparable-mass binary systems have neglected either
spin-precession effects
or subdominant harmonics and amplitude modulations. Here we give the first
explicit
description of how these effects combine to improve parameter estimation. We
consider supermassive black hole binaries as
expected to be
observed with the planned space-based interferometer LISA, and study the
measurement accuracy for several astrophysically interesting
parameters obtainable taking into account the full 2PN waveform for spinning
bodies, as well as spin-precession effects. We find that for binaries
with a total mass in the range $10^5 M_\odot < M < 10^7 M_\odot$ at a redshift
of $1$, a factor $\sim 1.5$
is in general gained in accuracy, with the notable exception of the
determination of the individual masses in equal-mass systems, for which a factor
$\sim 5$ can be gained. We also find, as could be expected, that using the full
waveform helps increasing the upper mass limit for detection, which can be as
high as $M = 10^8 M_\odot$ at a redshift of 1, as well as the redshift limit
where some information can be extracted from a system, which is roughly $z
\gtrsim 10$ for $M \leqslant 10^7 M_\odot$, $1.5 \upto 5$ times higher than with
the
restricted waveform. We computed that the full waveform allows to use
supermassive black hole binaries as standard sirens up to a redshift of $z
\approx 1.6$, about $0.4$ larger than what previous studies allowed. We
found that for lower unequal-mass binary systems, the measurement accuracy is
not
as drastically improved as for other systems. This suggests that for these
systems, adding parameters such as eccentricity or alternative gravity
parameters could be achieved without much loss in the accuracy.
\end{abstract}

\pacs{}

\maketitle


\section{Introduction}

Gravitational waves (GW's), once their observation is made possible, will
provide new
means of observing the Universe. So far, the vast majority of observations have
been made through
electromagnetic radiation, and gravitational waves will surely make visible
different aspects of the Universe. For example, one could probe the different
galaxy formation models by detecting supermassive black hole mergers in a
large redshift range~\cite{svh}. Or, as GW's
provide a good way of measuring the luminosity distance to their source, one
could combine gravitational and electromagnetic observations to build a robust
measurement of the Hubble diagram, which would be of
great interest for cosmology~\cite{schutz,holzhughes}.
Another possibility would be to measure alternative gravity
parameters~\cite{bbw,arunwill,stavridiswill}. This can potentially be a powerful
way to constrain such theories, as each observed GW will give an
independent measurement of their parameters.

The new generation of ground-based detectors, such as advanced LIGO, and the
space-based detector LISA
will probably make the direct detection of gravitational waves
possible. Some of the most important sources of gravitational waves are
the compact
binary systems, i.e. systems of two compact objects (white dwarfs, neutron
stars,
or black holes).
As such detections rely on matched filtering techniques, several groups have
made efforts in building accurate templates based on the post-Newtonian (PN)
approximation (see e.g.~\cite{th,oto,fbb,fbb2,abiq,abfo}). The limitations of
such results have then been checked to estimate the precision with
which one could measure
the properties of a system emitting such waves, for example its distance from
the Solar System, location in the sky, or the individual
masses of the objects forming it~\cite{finn, cutler, hughes, vecchio,
langhughes, aissv, portercornish, triassintes}.

 A binary system of compact objects emits gravitational waves during
three distinct phases, called inspiral, merger, and ringdown. During the
inspiral phase, most of the gravitational radiation is emitted at twice the
orbital frequency of the system, which slowly increases as it loses energy
emitting gravitational waves. As its two members come closer, higher
harmonics become more and more important, and more power gets emitted.
Finally, the two members enter the merger phase, where they have come so close
that they cannot be treated anymore as two separate objects, and begin to merge,
emitting complicated gravitational radiation that has not yet been described
but numerically. After that, the remnant begins to radiate away its
energy during the ringdown phase, where it approaches exponentially
the structure of a Kerr black hole. 
 
LISA has been designed to be particularly sensitive to binaries that contain a
supermassive black hole (SMBH, of mass $10^5 \upto 10^8 M_{\odot}$), which can
be separated into two (or three) categories, according to the companion mass:
extreme mass-ratio inspirals
(EMRIs) and supermassive black hole binaries (SMBHBs). A third category is
sometimes added: intermediate mass ratio inspirals (IMRIs), which somehow lies
between the two others.

EMRIs are inspirals with a mass ratio between the two members of the order of
$10^{-4} \upto 10^{-7}$, and are most accurately described by black hole
perturbation theory. Such events are likely to occur in the center of galaxies,
the majority of which are believed to host a SMBH, when a compact object of
stellar mass is ``eaten up'' by the central object. Several groups have attacked
the
problem of describing the form of the gravitational radiation emitted by such
objects~\cite{ckp,barackcutler,bfggh}.

SMBHBs are binaries containing two SMBHs, forming during the merging of two
galaxies, when they both host one in their center. Such events should be
much more rare than EMRIs, but some galaxy formation models~\cite{berti} and
observation of
nearby
galaxies~\cite{narayan} suggest that they might happen often enough to be
observable. As
these events are much louder than any other source, we could observe them at
very high redshifts (up to $z = 20$ or even higher), and thus tightly constrain
galaxy formation models.

The first attempt to estimate with which accuracy an interferometer could
measure the properties of a compact object binary was made by Finn~\cite{finn},
who first introduced the Fisher matrix analysis, which is now
widely used in this context. A few years later, Cutler~\cite{cutler} applied
this formalism
to LISA, focusing on the angular resolution that the space-based detector could
get for black hole binaries, using the Newtonian quadrupole formula.
Hughes~\cite{hughes} repeated the study including the PN expansion for
the frequency of the wave. Vecchio~\cite{vecchio}, then, considered
the case of the ``simple precession''~\cite{acst} of the angular momenta for
spinning BH's. Lang
and Hughes~\cite{langhughes} then used the full precession equations to further
refine the parameter estimation. Recently, Arun et al.~\cite{aissv} and
Porter and Cornish~\cite{portercornish}
included the full post-Newtonian waveform in the
context of nonspinning black holes, and Trias
and Sintes~\cite{triassintes} used it for spinning black holes neglecting
spin-precession effects. The LISA Parameter Estimation Taskforce~\cite{petf}
used the full waveform with spin-precession effects, without publishing a
detailed study of the expected statistical errors.
It is worth noting that all of the effects
that these works studied, as more and more precise waveforms were used, helped
to
improve subsequently the expected measurement accuracy of LISA. In this paper,
we study the question of whether the
inclusion of the full post-Newtonian waveform in the context of spinning black
holes undergoing spin-orbit precession helps to break more degeneracies, thus
helping to
further increase the accuracy in the measurement of the source parameters.

In Sec.~\ref{sec:theory}, we derive the waveform used to perform
our study, and we quote some basics of Fisher matrix analysis. In
Sec.~\ref{sec:simulations}, we describe the simulations that we ran. In
Sec.~\ref{sec:results}, we give the results of our simulations, and analyze
them from an astrophysical point of view. We conclude in
Sec.~\ref{sec:conclusion}.

\section{Theory}

 \label{sec:theory}

\subsection{Evolution}

The state of a binary system of two Kerr black holes at a given time in the
center of mass frame is fully described by 14 intrinsic parameters. These reduce
to 12 if we assume that the binary lies on a circular orbit. One possible choice
is to take as intrinsic parameters a unit vector pointing in the direction of
the orbital
angular momentum $\uvec{L}$, the orbital angular frequency $\omega$ (we will
reserve the symbol $f$ for arguments of Fourier transforms, and will express
the orbital frequency always with the angular frequency to avoid confusions),
the
individual spins of each black hole, $\bm{S}_1$ and $\bm{S}_2$,
their masses $m_1$ and $m_2$, and the orbital phase $\varphi$. To these
intrinsic parameters, we have to add three more extrinsic parameters which
locate the
binary in space. Those can be chosen to be $\uvec{n}$, a unit vector pointing in
the direction of the binary as seen from the Solar System, and $d_L$, the
luminosity distance from the binary to the Sun. We will denote all
unit vectors with a hat throughout this paper.

Another extrinsic parameter, the redshift, also plays a role in the
determination of the waveform, but it cannot be detected by GW observations.
Indeed, the redshift causes the observed angular frequency $f_o$ of the
wave to
decrease with
respect to the emitted one $f_e$, as $f_o = f_e / (1 + z)$. But
(see
following derivation) the exact same wave, within the post-Newtonian framework,
is emitted by a second system with parameters $m_i^{(2)} = (1 + z) m_i^{(1)}$,
$d_L^{(2)} = (1 + z) d_L^{(1)}$, not experiencing any redshift. Therefore, the
redshift and luminosity distance cannot be measured separately with a
gravitational wave observation, so that we have to assume a relation between the
two parameters. This implies that
observations of a light signal emitted during a merger, the
redshift of which is possible to determine, are of great astrophysical
interest.
During the whole derivation of the gravitational wave signal below, we
assume that the source is at redshift $z = 0$. The actual observed
wave can then be
easily determined redshifting the masses and luminosity distance, as we did
in our simulations.

To compute the relation between redshift and luminosity distance, we assume a
flat $\Lambda$CDM cosmology without radiation with the latest WMAP
parameters~\cite{wmap5}: $\Omega_\Lambda = 0.72$, $\Omega_m = 0.28$, $H_0 =
70.1$ km/s/Mpc.

The relation is then given by
\begin{equation}
 d_L(z) = (1+z) \frac{c}{H_0} \int_0^z \frac{dz'}{\sqrt{\Omega_m (1+z')^3 +
\Omega_\Lambda}}, \label{redlumdist}
\end{equation}
which can be determined numerically.

The problem of the motion of the system during the inspiral phase in
full General Relativity has been too hard to be solved so far. However, a great
effort has been made to attack the problem in the framework of the
post-Newtonian
formalism. The current state-of-the-art evolution equations go up to the 2.5PN
order beyond leading order for spinning objects~\cite{fbb}. As 2.5PN spin-orbit
and spin-spin
coupling terms are not yet known in the waveform, we chose to stop at the 2PN
level, up to which both the evolution equations and the waveform are known. We
will
use the following mass parameters for the derivation of the evolution equations
and of the waveform: the total mass $M = m_1 +
m_2$, the reduced mass $\mu = m_1 m_2/M$, and the symmetric mass ratio
$\nu = \mu/M$.

The 2PN orbit-averaged relation between the orbital angular frequency $\omega$
and the orbital
separation in harmonic coordinates $r$ is given by~\cite{fbb2}
\begin{multline} \label{om_of_gam}
 \omega = \frac{c^3}{GM} \gamma^{3/2} \bigg[ 1 + \left( \frac{\nu}{2} -
\frac{3}{2} \right) \gamma - \frac{1}{2} \beta(2,3) \gamma^{3/2} \\
 + \left( \frac{15}{8} + \frac{47\nu}{8} + \frac{3\nu^2}{8} - \frac{3}{4}
\sigma(1,3) \right) \gamma^2 \bigg],
\end{multline}
where the orbital separation parameter
$\gamma$ and the spin-orbit and spin-spin
couplings $\beta$ and $\sigma$ are given by
\begin{align}
 \gamma &\equiv \frac{GM}{rc^2}, \\
 \beta(a,b) &\equiv \frac{c}{G} \sum_{i=1}^2 \left( \frac{a}{M^2} +
\frac{b\nu}{m_i^{\phantom{i}2}} \right) \bm{S}_i \cdot \uvec{L}, \\
 \sigma(a,b) &\equiv \frac{c^2}{\nu M^4 G^2} \left( a \bm{S}_1 \cdot \bm{S}_2
- b \left( \bm{S}_1 \cdot \uvec{L} \right) \left( \bm{S}_2 \cdot
\uvec{L} \right) \right).
\end{align}

The evolution equation of the angular frequency is given at 2PN order
by~\cite{fbb2}
\begin{align}
 \frac{dx}{dt} &= \frac{64\nu}{5} \frac{c^3}{GM} x^{5}
\Bigg[ 1 - \left( \frac{743}{336} + \frac{11\nu}{4} \right) x \nonumber\\
 &+ \left( 4\pi -
\frac{1}{12} \beta(113,75) \right) x^{3/2} \label{domdt_of_x}\\
 &+ \left( \frac{34103}{18144} + \frac{13661\nu}{2016} + \frac{59\nu^2}{18} -
\frac{1}{48} \sigma(247,721)  \right) x^2 \Bigg], \nonumber
\end{align}
where $x$ is a dimensionless orbital frequency parameter defined as
\begin{equation}
 x \equiv \left( \frac{GM\omega}{c^3} \right)^{2/3}.
\end{equation} 

We can integrate Eq.~\eqref{domdt_of_x} to get
\begin{align}
 t &= t_c - \frac{5GM}{256 \nu c^3} x^{-4} \bigg[ 1 + \left( \frac{743}{252} +
\frac{11\nu}{3} \right) x \nonumber\\
&+ \left( \frac{2}{15} \beta(113,75) -
\frac{32}{5}\pi
\right) x^{3/2} \label{t_of_x}\\
 &+ \left( \frac{3058673}{508032} + \frac{5429\nu}{504} + \frac{617\nu^2}{72} +
\frac{1}{24} \sigma(247,721) \right) x^2 \bigg]. \nonumber
\end{align}

Integrating once more yields the orbital phase $\varphi = \int
\omega dt$, as a function of the orbital frequency parameter
\begin{align}
 \varphi(x) &= \varphi_c - \frac{x^{-5/2}}{32\nu} \bigg[ 1 +
\left(\frac{3715}{1008} + \frac{55\nu}{12} \right) x \nonumber\\
&\phantom{=}+ \left( \frac{5}{24}
\beta(113,75) - 10\pi \right) x^{3/2} \nonumber\\
&\phantom{=}+ \bigg(
\frac{15293365}{1016064} + \frac{27145\nu}{1008} \nonumber\\
&\phantom{=}+ \frac{3085\nu^2}{144} + \frac{5}{48} \sigma(247,721) \bigg) x^2
\bigg]. \label{phi_of_x}
\end{align}

The dragging of inertial frames induces a coupling between the individual spins
and the orbital angular momentum. The orbit-averaged conservative part of the
evolution
equations (i.e. without radiation reaction, $\dvec{L} + \dvec{S}_1 +
\dvec{S}_2 = 0$) are given for circular orbits at 2PN
order by
\begin{align}
 \dvec{L} &= \frac{G}{c^2} \frac{1}{r^3} \left( \left( 2 + \frac{3
m_2}{2
m_1} \right) \bm{S}_1 + \left( 2 + \frac{3 m_1}{2 m_2} \right) \bm{S}_2
\right) \times \bm{L} \nonumber\\
 &\phantom{=} - \frac{3 G}{2 c^2} \frac{1}{r^3} \left( \left( \bm{S}_2 \cdot
\uvec{L} \right) \bm{S}_1 + \left( \bm{S}_1 \cdot \uvec{L} \right)
\bm{S}_2
\right) \times \uvec{L}, \label{Lhatdot}\\
 \dvec{S}_i &= \frac{G}{c^2} \frac{1}{r^3} \left[ \left( 2 + \frac{3
m_j}{2 m_i} \right) \bm{L} + \frac{1}{2} \bm{S}_j - \frac{3}{2} \left(
\bm{S}_j \cdot \uvec{L} \right) \uvec{L} \right] \times \bm{S}_i,
\end{align}
where it is understood that $i \neq j$, $i,j \in \{1,2\}$, and the orbital
separation $r$ and the
norm of the orbital angular momentum $L$ are related to the orbital frequency by
their Newtonian relation:
\begin{align}
 L &= \mu \left( \frac{G^2M^2}{\omega} \right)^{1/3}, \\
 r &= \left( \frac{GM}{\omega^2} \right)^{1/3}.
\end{align}
Higher order relations would give corrections which exceed the 2PN order.

Using the above relations together with the first order of
Eq.~\eqref{domdt_of_x}, we
can change variables from time to orbital angular frequency,
and use the relations to express the precession equations:
\begin{align}
 \frac{d\bm{S}_i}{d\omega} &= \frac{5}{96} \frac{c^3}{G M} \omega^{-2}
\Bigg[
\uvec{L} \times \bm{\Sigma}_i  \nonumber \\
&+ \frac{1}{2L} \left(
\bm{S}_j - 3 \left(
\bm{S}_j \cdot \uvec{L} \right) \uvec{L} \right) \times \bm{S}_i
\Bigg],
\\
 \frac{d\uvec{L}}{d\omega} &= \frac{5}{96} \frac{c^3}{G M} \omega^{-2}
\frac{1}{L}\left[ \bm{\Sigma}_1 + \bm{\Sigma}_2 - \frac{3}{2L} \left(
\bm{\sigma}_1 + \bm{\sigma}_2 \right) \right] \times \uvec{L}
\label{dLhat_df}\\
 &= -\frac{1}{L} \left( \frac{d\bm{S}_1}{d\omega} +
\frac{d\bm{S}_2}{d\omega}
\right),
\nonumber
\end{align}
where
\begin{align}
 \bm{\Sigma}_i &= \left( 2 + \frac{3 m_j}{2 m_i} \right) \bm{S}_i, \\
 \bm{\sigma}_i &= \left( \bm{S}_j \cdot \uvec{L} \right) \bm{S}_i.
\end{align}

\subsection{Waveform}

The general form of a gravitational wave emitted by a two-body system, even
nonspinning, is not known
in the context of full general relativity. However, it has been computed in the
post-Newtonian framework. The results for a nonspinning binary system are
available at 2.5PN order~\cite{abiq}, and the spin effects at 2PN
order~\cite{abfo}.

A convenient way to define the phase of the wave observed in a detector is in
terms of the ``principal+ direction''~\cite{acst},
which is defined as the direction of the vector $\uvec{L} \times \uvec{n}$. As
the orbital angular momentum precesses, the principal+ direction changes, and
this must be taken into account in the waveform. This
effect amounts, at 2PN order, to
\begin{align}
 \delta \varphi &= - \int_t^{t_c} \frac{\uvec{L} \cdot \uvec{n}}{1 - \left(
\uvec{L} \cdot \uvec{n} \right)^2} \left( \uvec{L} \times \uvec{n} \right) \cdot
\duvec{L} dt \nonumber\\
 &= \delta\varphi_0 + \int_{\omega_0}^\omega \frac{\uvec{L} \cdot \uvec{n}}{1 -
\left(
\uvec{L} \cdot \uvec{n} \right)^2} \left( \uvec{L} \times \uvec{n} \right) \cdot
\frac{d\uvec{L}}{d\omega} d\omega, \label{deltaphiprec}
\end{align} 
where $\omega_0$ is an arbitrary constant corresponding to the time $t_0$,
$\delta \varphi_0 = - \int_{t_0}^{t_c}
(d\delta
\varphi/dt) dt$, and $d\uvec{L}/d\omega$ is given in Eq.~\eqref{dLhat_df}.

The 2PN accurate orbital phase is then given in terms of orbital
angular frequency by:
$\phi(\omega) = \varphi(\omega) + \delta \varphi(\omega)$.

The waveform is a series of harmonics of the orbital frequency:
\begin{equation}
 h_{+,\times} = \frac{2GM\nu x}{d_L c^2} \left[ \sum_{n \geqslant 0}
\left(A_{+,\times}^{(n)} \cos n\phi + B_{+,\times}^{(n)} \sin n\phi \right)
\right].
\end{equation}

The coefficients of the series take the form of post-Newtonian series:
\begin{align}
  A_{+,\times}^{(n)} = \sum_{i \geqslant 0} a^{(n,i/2)}_{+,\times} x^{i/2}, \\
  B_{+,\times}^{(n)} = \sum_{i \geqslant 0} b^{(n,i/2)}_{+,\times} x^{i/2}.
\end{align}

The exact form of the coefficients for a nonspinning system can be found
in~\cite{abiq,abfo}. Note, however, that both express their final result using
another phase which differs
from the orbital phase at 1.5PN order: $\Psi = \phi - 2
\log(\omega/\bar{\omega})
x^{3/2}$, where $\bar{\omega}$ is an arbitrary constant.
We put together the results from these two papers to build a coherent 2PN
accurate waveform for spinning bodies, see the
appendix.

\subsubsection{Extrinsic effects}
\label{exteff}

\begin{figure*}[!ht]
 \begin{center}
  \includegraphics[width=12cm]{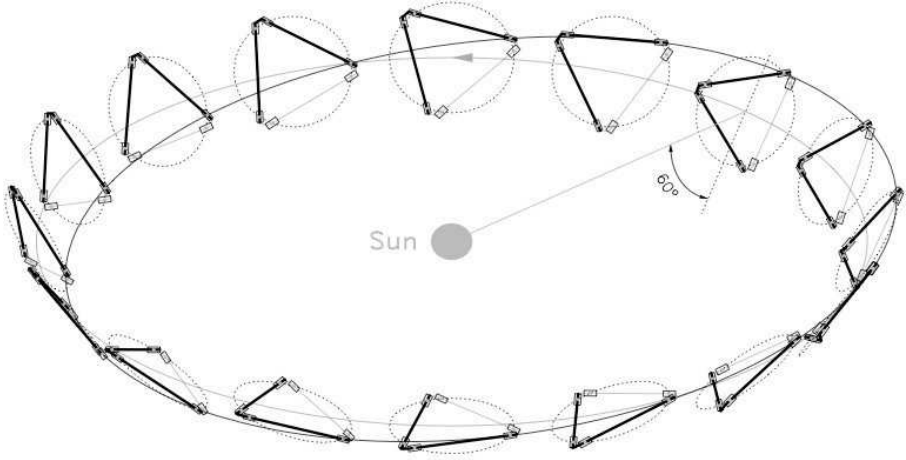}
  \caption{The orbit of LISA around the Sun as currently planned. Image taken
  from the LISA pre-phase-A report~\cite{LISAPPA}.\label{fig_LISA_orbit}}
 \end{center}
\end{figure*}

The LISA constellation will consist in three spacecrafts launched in orbit
around the Sun, at a mean distance of 1~AU, on slightly eccentric orbit so that
the spacecrafts stay at the same distance from each other all along the year.
The barycenter of LISA will be located on the orbit of the Earth, 20$^\circ$
behind it, and the normal to the plane on which the spacecrafts lie will make a
60$^\circ$ angle with the normal to the ecliptic, see
Fig.~\ref{fig_LISA_orbit}.

To describe extrinsic effects that depend on the position of LISA, we
follow~\cite{cutler} and define two different frames: a frame tied to the
detector,
$(x,y,z)$, and a fixed, Solar System frame, tied to the distant stars
$(\bar{x},\bar{y},\bar{z})$ (we consider that the motion of the Sun with
respect to the distant stars can be neglected during the lifetime of the LISA
mission).

The unit vectors along the arms of LISA $\uvec{l}_i$, $i \in \{1,2,3\}$ are
defined
in the detector frame:
\begin{align}
 \uvec{l}_i &= \cos \gamma_i \uvec{x} + \sin \gamma_i \uvec{y}, \nonumber\\
 \gamma_i &= \frac{\pi}{12} + (i-1) \frac{\pi}{3}.
\end{align} 

The $(\bar{x},\bar{y})$ plane of the Solar System frame is defined to be the
ecliptic, so that the spherical angles of the barycenter of LISA are
\begin{equation}
 \bar{\Theta} = \frac{\pi}{2}, \qquad \bar{\Phi}(t) = 2\pi t/T,
\end{equation}
where $T = 1$ yr, and we chose that $\bar{\Phi} = 0$ at $t=0$.

The waveform is given relative to the Solar System frame. To take into account
the fact that the detector is not static in this frame, we have to add a phase
to each harmonic, which is equivalent to add the so-called
Doppler phase to the orbital phase:
\begin{equation}
 \phi_D(t) = \frac{\omega R}{c} \sin \bar{\theta}_N \cos (\bar{\Phi}(t) -
\bar{\phi}_N),
\end{equation} 
where $R=1$~AU, and $\bar{\theta}_N$ and $\bar{\phi}_N$ are the spherical
angles of
the position of the source in the Solar System frame.

The orbital phase then becomes
\begin{equation}
 \psi = \phi + \phi_D = \varphi + \delta \varphi + \phi_D
\end{equation}

The normal to the detector plane $\uvec{z}$ is at constant angle $\bar{\theta}_z
= \pi/3$ from the normal to the ecliptic $\bm{\hat{\bar{z}}}$, and
constantly points
in the direction of the $\bar{z}$-axis from the barycenter of LISA. Furthermore,
each
satellite rotates around the $\uvec{z}$-axis once a year (see
Fig.~\ref{fig_LISA_orbit}). Let us
express then the detector frame in the Solar System frame, assuming that
$\uvec{y}
\cdot \bm{\hat{\bar{y}}} = 1$ at $t=0$:
\begin{align}
 \uvec{x} &= \left( \frac{3}{4} - \frac{1}{4} \cos 2 \bar{\Phi}(t) \right)
\bm{\hat{\bar{x}}} - \frac{1}{4} \sin 2 \bar{\Phi}(t)
\bm{\hat{\bar{y}}}\nonumber \\
&\qquad\qquad\qquad\qquad\qquad+
\frac{\sqrt{3}}{2} \cos \bar{\Phi}(t) \bm{\hat{\bar{z}}}, \\
 \uvec{y} &= - \frac{1}{4} \sin 2 \bar{\Phi}(t) \bm{\hat{\bar{x}}} + \left(
\frac{3}{4} + \frac{1}{4} \cos 2 \bar{\Phi}(t) \right) \bm{\hat{\bar{y}}} 
\nonumber\\
&\qquad\qquad\qquad\qquad\qquad+
\frac{\sqrt{3}}{2} \sin \bar{\Phi}(t) \bm{\hat{\bar{z}}}, \\
 \uvec{z} &= - \frac{\sqrt{3}}{2} \cos \bar{\Phi}(t) \bm{\hat{\bar{x}}} -
\frac{\sqrt{3}}{2} \sin \bar{\Phi}(t) \bm{\hat{\bar{y}}} + \frac{1}{2}
\bm{\hat{\bar{z}}}.
\end{align}

LISA will act during the incoming of a gravitational wave as a pair of two-arm
detectors, but with a response scaled by a $\sqrt{3}/2$ factor due to the
$60^\circ$ opening angle of the constellation, following
the pattern ($k = 1,2$):
\begin{align}
 h_k &= \frac{\sqrt{3}}{2} \left( F_{k}^{+} h_+ + F_{k}^{\times} h_\times
\right), \\
 F_1^+(\theta_N, \phi_N, \psi_N) &= \frac{1}{2} \left( 1 + \cos^2 \theta_N
\right) \cos 2\phi_N \cos
2\psi_N \nonumber\\
&\phantom{=}- \cos \theta_N \sin 2\phi_N \sin 2\psi_N,\\
 F_1^\times(\theta_N, \phi_N, \psi_N) &= F_1^+(\theta_N,\phi_N,\psi_N-\pi/4),\\
 F_2^+(\theta_N, \phi_N, \psi_N) &= F_1^+(\theta_N,\phi_N-\pi/4,\psi_N),\\
 F_2^\times(\theta_N, \phi_N, \psi_N) &=
F_1^+(\theta_N,\phi_N-\pi/4,\psi_N-\pi/4),
\end{align}
where $\theta_N$ and $\phi_N$ are the spherical angles of the position of the
binary
in the detector frame, and $\psi_N$ is defined through
\begin{equation}
 \tan \psi_N \equiv \frac{\uvec{L}\cdot\uvec{z} - (\uvec{L}\cdot\uvec{n})
(\uvec{z}\cdot\uvec{n})}{\uvec{n}\cdot( \uvec{L}\times\uvec{z} )}.
\end{equation}

We expressed here two combinations of the response of the three arms of LISA
whose detector noises are uncorrelated~\cite{cutler}.

Using this, we find the response function of the detectors ($k = 1,2$):
\begin{align}
 h_k &= \frac{\sqrt{3}GM\nu x}{d_L c^2} \sum_{n \geqslant 0} \Bigg[ \nonumber \\
 &\phantom{=} \sum_{i \geqslant 0}
\left(F_k^+(t)
a_+^{(n,i/2)} + F_k^\times(t) a_\times^{(n,i/2)}\right)x^{i/2} \cos n\psi
\nonumber\\
 &\phantom{=}+ \sum_{i \geqslant 0} \left(F_k^+(t) b_+^{(n,i/2)} +
F_k^\times(t) b_\times^{(n,i/2)}\right)x^{i/2} \sin n\psi 
\Bigg] \\
 &= \frac{\sqrt{3}GM\nu x}{d_L c^2} \sum_{n \geq 0} \left[
A_{k,n}(t) \cos n\psi + B_{k,n}(t) \sin n \psi \right].
\end{align} 

We can change this into the phase-amplitude representation:
\begin{multline}
 h_k = \frac{\sqrt{3}GM\nu x}{d_L c^2} \bigg[ A_+^{(0)}(t) F_k^+(t) \\
 + \sum_{n \geq 1}
A_{k,n}^{pol}(t) \cos\left( n \psi + \phi_{k,n}^{pol}(t) \right) \bigg],
\end{multline} 
where $\phi_{k,n}^{pol}$ is the polarization phase, and $A_{k,n}^{pol}$ is the
polarization amplitude:
\begin{align}
 \tan \phi_{k,n}^{pol} &= - \frac{B_{k,n}}{A_{k,n}}, \label{phipol}\\
 A_{k,n}^{pol} &= \mathrm{sgn}(A_{k,n}) \sqrt{A_{k,n}^2 + B_{k,n}^2}.
\label{Apol}
\end{align}

The final form of the gravitational wave signal is thus
\begin{align}
 h_k &= \frac{\sqrt{3}GM\nu x}{d_L c^2} \left[ A_+^{(0)} F_k^+ + \sum_{n
\geqslant 1}
A_{k,n}^{pol}
\cos\psi_{k,n} \right], \\
 \psi_{k,n} &= n (\varphi + \delta\varphi + \phi_D) + \phi_{k,n}^{pol}.
\end{align}

To estimate the measurement error in the different parameters of the binary, we
 need to know the Fourier transform of the signal
$\tilde{h}_k(f)$~\footnote{Note
that the symbol $f$ here and in the following denotes the argument of the
Fourier transform of the signal, and not the orbital frequency.}.
\begin{align}
 \tilde{h}_k(f) &= \int_{-\infty}^\infty h_k(t) e^{2\pi i f t} dt \nonumber\\
 &= \frac{\sqrt{3}GM\nu}{d_L c^2} \Bigg[ \int_{-\infty}^\infty x \sum_{n
\geqslant 1}
A_{k,n}^{pol}
\cos\psi_{k,n} e^{2\pi i f t} dt \nonumber\\
 &\phantom{=}+ \int_{-\infty}^\infty x A_+^{(0)} F_k^+
e^{2\pi i f t} dt
\Bigg] \nonumber\\
 &\approx \frac{\sqrt{3}GM\nu}{2 d_L c^2} \sum_{n \geqslant 1} \Bigg[
\int_{-\infty}^\infty
x A_{k,n}^{pol}
e^{i (2\pi ft + \psi_{k,n})} dt \nonumber\\
&\phantom{=}+ \int_{-\infty}^\infty x A_{k,n}^{pol} e^{i
(2\pi ft -
\psi_{k,n})} dt \Bigg].  \label{fthk}
\end{align}

Note that we neglected in the last line the Fourier transform of the so-called
memory effect $A_+^{(0)}$. This is based on the fact that the Fourier transform
of the
function $x A_+^{(0)} F_k^+$ accumulates around frequencies which are separated
from
the
orbital frequency range, at least during most of the inspiral. It will thus not
contribute to the relevant frequencies.

To compute the integrals, we rely on the stationary phase approximation.
Neglecting the integrals with the $e^{i (2\pi ft + \psi_{k,n})}$ factor, as they
will only contribute to negative frequencies, the stationary points for the
other
integrals are given by
\begin{equation}
 2 \pi f = \psi_{k,n}'(t_{k,n}) = n \omega(t_{k,n}) + n \phi_D'(t_{k,n}) +
(\phi_{k,n}^{pol})'(t_{k,n}).
\end{equation} 

For the same reasons as before, we can safely neglect the derivatives of the
Doppler phase and of the polarization phase. We thus get the following
expression for
the stationary point:
\begin{equation}
 t_{k,n} = t_n = t(f/n),
\end{equation} 
where the function $t(f)$ is defined at 2PN order by Eq.~\eqref{t_of_x}.

Thus, we get the following expression for the Fourier transform of the
gravitational wave signal:
\begin{align}
 \tilde{h}_k(f) &= \frac{\sqrt{5\pi\nu}G^2M^2}{8d_L c^5} \sum_{n \geqslant 1}
A_{k,n}^{pol}[t(f/n)] x_n^{\phantom{n}-7/4} S(f/n) \nonumber\\
 &\phantom{=} \cdot \exp\bigg\{i \Big[ n\Big(\Psi(f/n) - \delta\varphi(f/n)
\label{htilde} \\
 &\qquad\qquad\qquad-
\phi_D[t(f/n)] \Big) - \phi_{k,n}^{pol}[t(f/n)] \Big] \bigg\}, \nonumber
\end{align} 
where $x_n = x(f/n) = n^{-2/3}x$, and
\begin{align}
 S(f) &= \Bigg[ 1 + \left( \frac{743}{672} + \frac{11\nu}{8} \right) x
\nonumber\\
 &\qquad\qquad + \left(
\frac{1}{24} \beta(113,75) - 2\pi) \right) x^{3/2} \nonumber\\
 &\qquad\qquad + \bigg( \frac{7266251}{8128512} + \frac{18913\nu}{16128}
\nonumber\\
 &\qquad\qquad+
\frac{1379\nu^2}{1152} + \frac{1}{96} \sigma(247,721) \bigg) x^2 \Bigg], \\
 \Psi(f) &= \left( \frac{t_c c^3}{GM} \right) x^{3/2} - \varphi_c
-\frac{\pi}{4} \nonumber\\
 &\qquad\qquad+ \frac{3 x^{-5/2}}{256\nu} \Bigg[ 1 + \left( \frac{3715}{756} +
\frac{55\nu}{9} \right) x \nonumber\\
&\qquad\qquad+ \left( \frac{1}{3} \beta(113,75) - 16\pi
\right)
x^{3/2} \nonumber\\
 &\qquad\qquad + \bigg( \frac{15293365}{508032} + \frac{27145 \nu}{504}
\nonumber\\
 &\qquad\qquad+ 
\frac{3085 \nu^2}{72} + \frac{5}{24} \sigma(247,721) \bigg) x^2 \Bigg].
\end{align}
where we used here $x = x(\omega = 2\pi f)$ a different orbital frequency
parameter for each harmonic.

Note that Lang and Hughes~\cite{langhughes} took the zeroth order form for $S$,
$S(f) = 1$, consistently with neglecting all amplitude modulations.

Finally, a binary will be observed with LISA during a finite amount of time.
Therefore, if we denote by $t_i$ and $t_f$ respectively the initial and final
time of observation, the orbital frequencies available for the Fourier transform
will lie between $f_{orb}(t_i)$ and $f_{orb}(t_f)$. Thus, the final Fourier
transform will be of the form:
\begin{equation}
 \tilde{h}_k(f) = \sum_{n \geqslant 1} \tilde{h}_{k,n}(f) \theta(f -
nf_{orb}(t_i))
 \theta(nf_{orb}(t_f) - f),
\end{equation} 
where $\theta$ is the Heaviside step function.

We compared in this work three different waveforms, which we called full
waveform (FWF), simplified waveform (SWF), and restricted waveform (RWF). The
latter is the one used in~\cite{langhughes}.

The FWF contains all post-Newtonian corrections of the frequency and
amplitude of the wave up to 2PN order. It is obtained using the amplitudes
given in the
appendix in Eqs.~\eqref{phipol} and~\eqref{Apol},
and inserting the results in Eq.~\eqref{htilde}.

The SWF contains all post-Newtonian corrections of the
frequency up to 2PN order, and the lowest order amplitude of each harmonic
present at the 2PN level. With this approximation, we find particularly
simple forms for the polarization amplitudes and phases (with $F_{+,\times} =
F_k^{+,\times}$, $c_i = \uvec{L} \cdot \uvec{n}$, $s_i = |\uvec{L}
\times \uvec{n}|$):
\begin{subequations}
\begin{align}
 A_{k,1}^{pol} &= - \mathrm{sgn}(F_+) \frac{x^{1/2} s_i}{8} \sqrt{1-4\nu} \
\cdot \nonumber\\
&\qquad\qquad\qquad \sqrt{F_+^2 \left(5  +  c_i^2\right)^2  + 36 F_\times^2
c_i^2}, \\
 A_{k,2}^{pol} &= - \mathrm{sgn}(F_+) \sqrt{F_+^2\left(1 + c_i^2\right)^2 + 4
F_\times^2 c_i^2}, \\
 A_{k,3}^{pol} &= -\frac{9 x^{1/2} s_i}{8} \sqrt{1-4\nu} A_{k,2}^{pol},\\
 A_{k,4}^{pol} &= \frac{4 x s_i^2}{3} (1-3\nu) A_{k,2}^{pol},\\
 A_{k,5}^{pol} &= - \frac{625 x^{3/2} s_i^3}{384} \sqrt{1-4\nu} (1-2\nu)
A_{k,2}^{pol},\\
 A_{k,6}^{pol} &= \frac{81 x^2 s_i^4}{40} (1 - 5\nu + 5\nu^2) A_{k,2}^{pol},\\
 \phi_{k,1}^{pol} &= - \arctan\left( \frac{6 c_i F_\times}{(5 + c_i^2) F_+}
\right), \\
 \phi_{k,n}^{pol} &= - \arctan\left( \frac{2 c_i F_\times}{(1 + c_i^2) F_+}
\right), \qquad n \geqslant 2.
\end{align}
\end{subequations}
The SWF is obtained inserting the polarization amplitudes and phases above into
Eq.~\eqref{htilde} and, consistently with neglecting all amplitude
corrections, taking the lowest order of the overall amplitude correction $S(f) =
1$.

The RWF contains all post-Newtonian corrections of the
frequency up to 2PN order, and the lowest order amplitude of the second
harmonic. It is identical to the SWF, with the further approximation
$A^{pol}_{k,n} = \phi^{pol}_{k,n} = 0, \ n \neq 2$.

\subsection{Data analysis}

\label{sec:dataal}

The signal will of course also include noise.
A good description of the impact of noise can be found in~\cite{langhughes},
and we will refer to that study for how to model it. A deeper study of the
Fisher information formalism in the context of gravitational wave experiments
can be found in~\cite{vallisneri}.

As defined in~\cite{langhughes}, the inner product in the space of signals is
\begin{equation}
 (a|b) \equiv 4 \mathrm{Re} \int_0^\infty \frac{\tilde{a}^*(f)
\tilde{b}(f)}{S_h(f)}
df, \label{scalprod}
\end{equation} 
where $S_h(f)$ is the noise spectral density.

Now, if we have a signal $h$, described by a certain set of $n$ parameters
$\bm{\theta}$, the Fisher information matrix is an $n$-by-$n$ symmetric matrix
defined by
\begin{equation}
 \Gamma_{ij} \equiv \left( \der{h}{\theta^i} \right|\left. \der{h}{\theta^j}
\right).
\end{equation}
When we have several detectors, the Fisher information matrices are simply
added:
\begin{equation}
 \Gamma_{ij} = \sum_k \Gamma_{ij}^{(k)}.
\end{equation}
Finally, the covariance matrix is the inverse of the information matrix:
\begin{equation}
 \Sigma \equiv \Gamma^{-1}.
\end{equation} 

Its off-diagonal elements represent correlation coefficients between the
different parameters, and must satisfy
\begin{equation}
 \norm{\frac{\Sigma^{ij}}{\sqrt{\Sigma^{ii}\Sigma^{jj}}}} < 1,
\end{equation}
whereas its diagonal elements represent lower limits to
statistical errors on
their measurement:
\begin{equation}
 \Delta\theta^i = \sqrt{\Sigma^{ii}}.
\end{equation}

We chose to use a different noise curve than~\cite{langhughes}, motivated by the
fact that an agreement seems to have been found among the LISA parameter
estimation community~\cite{petf} to use a piecewise fit of the
expected noise, used in the second round of the mock LISA data
challenge~\cite{MLDC2}. We use the noise model described in~\cite{petf},
which consists in a sum of an instrumental noise $S_n(f)$ and a Galactic
confusion noise $S_{\mathrm{conf}}(f)$~\footnote{We found a typo in the
confusion noise, which made it discontinuous at $10^{-3}$~Hz. For $f \leqslant
10^{-3}$~Hz, we replaced $10^{-44.62} f^{-2/3}$ with the correct value of
$10^{-44.62} f^{-2.3}$.}.
The instrumental and confusion noise curves read ($f$ is given in Hertz)
\begin{subequations}
\begin{align}
 S_n(f) &= \frac{1}{L^2} \Bigg\{ \left[ 1 + \frac{1}{2} \left( \frac{f}{f_*}
\right)^2 \right] S_p \nonumber\\
 & \qquad\qquad+ \left[ 1 + \left( \frac{10^{-4}}{f} \right)^2
\right]
\frac{4 S_a}{\left(2\pi f\right)^4} \Bigg\}, \\
S_{\mathrm{conf}}(f) &= \left\{
\begin{array}{l l}
 10^{-44.62} f^{-2.3} & (f \leqslant 10^{-3}), \\
 10^{-50.92} f^{-4.4} & (10^{-3} < f \leqslant 10^{-2.7}), \\
 10^{-62.8} f^{-8.8} & (10^{-2.7} < f \leqslant 10^{-2.4}), \\
 10^{-89.68} f^{-20} & (10^{-2.4} < f \leqslant 10^{-2}), \\
 0 & (10^{-2} < f).
 \end{array}
  \right.
\end{align}
\end{subequations}
where $L = 5 \cdot 10^9~\mathrm{m}$ is the arm length of LISA, $S_p = 4 \cdot
10^{-22}~\mathrm{m}^2~\mathrm{Hz}^{-1}$ is the white position noise level, $S_a
= 9 \cdot 10^{-30}~\mathrm{m}^2~\mathrm{s}^{-4}~\mathrm{Hz}^{-1}$ is the white
acceleration noise level, and $f_* = c/(2\pi L)$ is the arm transfer frequency.
This way, we have $S_h(f) = S_n(f) + S_{\mathrm{conf}}(f)$.

\begin{figure}[!ht]
  \includegraphics[width=\columnwidth]{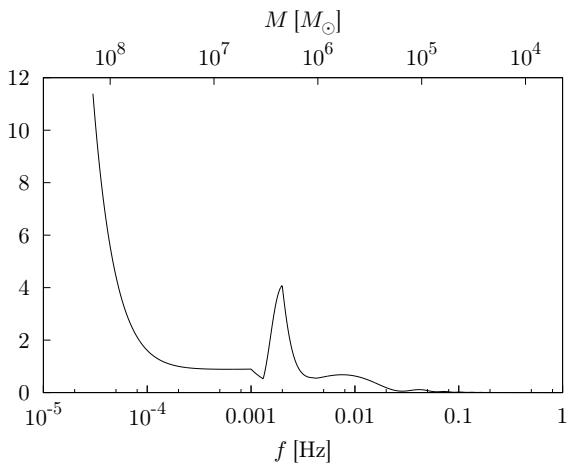}
  \caption{Ratio of the noise curve we used over the one
used by Lang and Hughes, as a function of the frequency. We added on top the
total mass of a binary which emits its second harmonic at the corresponding
maximum frequency. At high
frequencies ($f \gtrsim 5 \cdot 10^{-2}$~Hz), the curve used by Lang and Hughes
is inaccurate. There is
a peak at $f = 2 \cdot 10^{-3}$~Hz, where our noise is 4 times larger, and
which corresponds to the maximum frequency of binaries with a total mass $M
\approx 2 \cdot 10^6 M_\odot$. Between $10^{-3}$ and $2 \cdot 10^{-4}$~Hz, we
find good agreement, and for lower frequencies our noise is much
larger.\label{fig:noisecomp}}
\end{figure}

We found substantial differences in the noise curve that we used with respect
the one used by Lang and Hughes~\cite{langhughes}. We plot the ratio of the two
curves in Fig.~\ref{fig:noisecomp}.

Our noise is in general larger for frequencies below $10^{-2}$~Hz, which
corresponds to the maximum frequency of the RWF emitted by binaries with total
mass $M \approx 5
\cdot 10^5 M_\odot$ ($f_{max}$ is proportional to $1/M$). Therefore we expect
our
results to be more pessimistic than the ones in~\cite{langhughes}. However, we
do not expect the comparison between the different waveforms to depend very
much on the particular noise curve used in such a study.

We also compared the two instrumental noises and the two confusion
noises
separately. The two confusion noises are similar below $10^{-3}$~Hz and differ
above this value. They become negligible compared to the instrumental noise
above $5 \cdot 10^{-3}$~Hz. The peak visible in Fig.~\ref{fig:noisecomp} at
$f = 2 \cdot 10^{-3}$~Hz is
due to the differences in confusion noises. The discrepancies come from the
fact that after the publication of~\cite{langhughes}, a better estimation of
what fraction of low-mass binaries could be resolved and not contribute to the
confusion noise has been made~\cite{cornishlittenberg} and used in the noise
model of~\cite{petf}. The instrumental noise used in~\cite{langhughes} is a
low-frequency approximation based on the online sensitivity curve generator
provided by S.~Larson. The noise we used differs from the one used by Larson for
frequencies below $10^{-4}$~Hz. This comes from the fact that Larson assumes
white acceleration noise, whereas the authors of~\cite{petf}
additionally considered it to increase as $1/f$ below $10^{-4}$~Hz.

\section{Simulations}

\label{sec:simulations}

As a set of 12 intrinsic plus 3 extrinsic parameters for our simulations, we
used:
\begin{enumerate}
 \item[(i)] $\log_{10} m_1/M_\odot$ and $\log_{10} m_2/M_\odot$, for the 
masses of the two black holes.
 \item[(ii)] $\mu_l = \cos \theta_l$ and $\phi_l$, for the spherical angles of
the
orbital angular momentum $\bm{L}$ at $\gamma=\frac{1}{6}$.
 \item[(iii)] $\mu_1 = \cos \theta_1$ and $\phi_1$ for the spherical angles of
the
spin of the first black hole
$\bm{S}_1$ at $\gamma=\frac{1}{6}$.
 \item[(iv)] $\chi_1 = \frac{c}{Gm_1^2}
\norm{\bm{S}_1}$ for the dimensionless strength of the spin of the first
black hole, which has to satisfy $0
\leqslant
\chi_1 < 1$.
 \item[(v)] $\mu_2 = \cos \theta_2$, $\phi_2$, and $\chi_2$, same for the second
black hole as for the first one.
 \item[(vi)] $t_c$, the time of coalescence.
 \item[(vii)] $\varphi_c$, the phase at coalescence. As this phase is random and
its determination is not of any astrophysical interest, we can
safely neglect constants in the orbital phase, in particular $\delta\varphi_0$
from Eq.~\eqref{deltaphiprec}.
 \item[(viii)] $\mu_n=\cos\theta_n$ and $\phi_n$, the spherical angles of the
 position of the binary in the sky.
 \item[(ix)] $d_L$, the luminosity distance between the source and the Solar
System.
\end{enumerate}

All angles are taken in the frame tied to the distant stars.
We fix the zero point of time by the beginning of the LISA mission.

We performed Monte Carlo simulations where, given a set of parameters, we
evolved
the binary backwards in frequency starting from $\omega(\gamma=1/6)$ using a
fourth
order adaptive Runge-Kutta algorithm to find
$\uvec{L}(\omega)$, $\bm{S}_1(\omega)$, $\bm{S}_2(\omega)$, and
$\delta\varphi(\omega)$. We
stopped the simulations either at $t=0$, or when the frequency of the
highest harmonic had gone below the LISA band, for $6 \omega < 3
\cdot 10^{-5}~\mathrm{Hz}$. We chose to start at $\gamma=1/6$ because it is the
radius of the innermost stable circular orbit (ISCO) for a Schwartzschild black
hole. Of course, when dealing with a spinning black hole, the radius of the ISCO
can vary between $\gamma = 1$ and $\gamma = 1/9$, depending on the spin
parameter of the black hole and on the orientation of the orbit, but the series
may not converge when $\gamma$ is close to 1, so that we
chose to consider the post-Newtonian expansion as accurate for $\gamma
\leqslant 1/6$, which seems to be a good enough
prescription~\cite{bcp,bmtck}.
 
We then put these functions inside the Fourier transforms of the
waves~\eqref{htilde}.

Five derivatives $\partial_{\theta^i} \tilde{h}_k$ out of the 15 needed can be
found analytically. Three simple ones are:
\begin{align}
 \frac{\partial\tilde{h}_k(\theta^j,f)}{\partial t_c} &= 2 \pi i f
\tilde{h}_k(\theta^j,f), \\
 \frac{\partial\tilde{h}_k(\theta^j,f)}{\partial d_L} &= -
\frac{\tilde{h}_k(\theta^j,f)}{d_L}, \\
 \frac{\partial\tilde{h}_k(\theta^j,f)}{\partial \varphi_c} &= -i \sum_n n
\tilde{h}_{k,n}(\theta^j,f),
\end{align}
where $\tilde{h}_{k,n}$ is the $n$-th harmonic component of $\tilde{h}_k$. The
other two are the derivatives with respect to $\mu_n$ and $\phi_n$. The
only quantities in Eq.~\eqref{htilde} that depend on these parameters are
$A_{k,n}^{pol}$, $\phi_{k,n}^{pol}$, $\delta\varphi$, and $\phi_D$.

The derivatives which we could not find analytically have been computed
numerically using the relation
\begin{equation}
 \frac{\partial \tilde{h}_k (\theta^j,f)}{\partial \theta^i} \approx
\frac{\tilde{h}_k(\theta^j + \epsilon \delta^{ij} /2, f) - \tilde{h}_k(\theta^j
-\epsilon \delta^{ij}/2, f)}{\epsilon},
\end{equation}
where $\epsilon$ is a small displacement of the parameter $\theta^i$. We
used a constant value of $\epsilon = 10^{-7}$ for every parameter, except
for $\phi_l$ for which $\epsilon$ was divided by $2 - 2|\mu_l|$, $\mu_i$
($i \in \{1,2\}$)  for which $\epsilon$ was divided by $5 \chi_i$,
and $\phi_i$ for which $\epsilon$ was divided by $10 \chi_i(1 - |\mu_i|)$. The
formula is accurate up to $O(\epsilon^2)$.

We computed the functions $\uvec{L}(\omega)$, $\bm{S}_1(\omega)$,
$\bm{S}_2(\omega)$, and
$\delta\varphi(\omega)$ for each displacement of the parameters.

We then evaluated numerically the integrals
$(\partial_{\theta^i}\tilde{h}_k|\partial_{\theta^j}\tilde{h}_k)$ to find the
Fisher information matrix. Each harmonic $\tilde{h}_{k,n}(f)$
is truncated if necessary to remain inside the LISA band, which we take to be
$[3 \cdot 10^{-5}, 1]~\mathrm{Hz}$.

We added the contributions from both detectors, and then inverted the matrix
numerically to find the statistical error estimates.

We found that in some extreme situations, when $\uvec{L} \cdot \uvec{n}$ gets
close to 1, the Runge-Kutta method fails to converge when computing
$\delta\varphi$, because
\begin{equation}
 \frac{d\delta \varphi}{d\omega} = \frac{\uvec{L} \cdot \uvec{n}}{1 -
\left(
\uvec{L} \cdot \uvec{n} \right)^2} \left( \uvec{L} \times \uvec{n} \right) \cdot
\frac{d\uvec{L}}{d\omega} \sim \frac{1}{\norm{\uvec{L} \times \uvec{n}}}, \ 
\uvec{L} \cdot \uvec{n} \to 1.
\end{equation}
In those situations, we chose to compute $\delta\varphi(\omega)$ whenever
$\uvec{L} \cdot \uvec{n}$ is to close to 1 with an
approximate value,  which is
\begin{multline}
 \delta\varphi(\omega + \delta\omega) \approx \\
 \delta\varphi(\omega) +
\mbox{angle}
\left[ \left( \uvec{L}(\omega+\delta\omega) \times \uvec{n} \right) ,  \left(
\uvec{L}(\omega) \times \uvec{n} \right) \right].
\end{multline}

We ran different sets of simulations fixing the redshift and the masses, and
selected the other parameters randomly, using a flat distribution. The bounds to
put for the
Monte Carlo selection of the different parameters are clear, except for $t_c$.
We chose the following bounds, consistently with~\cite{langhughes}: the lower
bound for $t_c$ is for the physical separation parameter $\gamma$ to be equal
to $1/6$
at $t=0$ (combining Eqs.~\eqref{om_of_gam} and~\eqref{t_of_x}), and the
higher bound is $t_c = 3~\mathrm{yrs}$ (this is in fact the minimum science
requirement of the mission), which we take to be the duration
of the
LISA mission, so that the coalescence is visible during it.
We computed for each set the mean measurement errors for the parameters and
signal-to-noise ratio (SNR), comparing the output for the RWF, the SWF, and
the FWF defined at the end of Sec.~\ref{exteff}.

\section{Results}

\label{sec:results}

We ran different sets of simulations, each of them at a redshift of $z=1$,
varying the masses between $O(10^5 M_\odot)$ and $O(10^8 M_\odot)$, and the
mass ratio between 1:1 and 1:10.
We did not vary the
redshift, because, as described
in Sec.~\ref{sec:theory}, it cannot be measured by a gravitational wave
experiment. Furthermore, with a redshift to luminosity distance relation
fixed,
the only parameter varying with the 
redshift for constant redshifted masses is the luminosity distance. The
statistical errors in this case 
scale for all parameters as $(1+z)d_L$, as this parameter appears only as
an overall factor in the waveforms. For example, the
statistical error estimates on the parameters for a system with $m_1 = 5 \cdot
10^6 M_\odot$, $m_2 = 5 \cdot 10^5 M_\odot$, and $z=1$ are exactly the same as
those for
a system with $m_1 = 2 \cdot 10^6 M_\odot$, $m_2 = 2 \cdot 10^5 M_\odot$, and
$z=4$ (same redshifted masses), multiplied by $2.5 d_L(z=4)/d_L(z=1)$.
For binaries with masses higher
than $10^7 M_\odot$, the results for the RWF cannot be trusted, as the second
harmonic spends typically only a few orbits inside the LISA band, and even no
signal at all can be observed for $10^8 M_\odot$
binaries.
Each of our sets of simulations consisted in over a thousand binaries. We
performed \textit{a posteriori} statistical checks showing that the medians
should be
correctly estimated up to a few percent.

We present here in tables for all samples and all interesting parameters a
best-case measurement error ($5\%$ quantile), a typical error (the median), and
a worst-case error ($95\%$ quantile). The parameters we are
interested in are the (redshifted) individual masses of the black holes, shown
in Tables~\ref{tab:m1} and~\ref{tab:m2}  their
spin
parameters, shown in Tables~\ref{tab:ch1} and~\ref{tab:ch2}, the principal axes
of the localization ellipse in the sky, shown in Tables~\ref{tab:2a}
and~\ref{tab:2b}, and the
(redshifted)
luminosity distance to the source, shown in Table~\ref{tab:dl}.

We followed~\cite{langhughes} to present as sensible quantities for the sky
localization the principal axes $2a$ and $2b$ of the ellipse enclosing the
region outside of
which there is a $1/e$ probability of finding the binary.

For binaries for which no signal can be extracted from the data,
we fixed the errors to $\infty$. For errors apparently meaningless, such as
$\Delta d_L / d_L > 1$ or $2a > 2\pi$, we still provide the error as computed,
because it can give an indication on up to what redshift quantities can be
computed using the scaling property of the error with respect to $(1+z) d_L$.

\begin{table*}[!ht]
\begin{tabular}{|c|c|c|c|c|c|c|c|c|c|c|}
 \hline
 $m_1 [M_\odot]$ & $m_2 [M_\odot]$ & \multicolumn{9}{c|}{$\Delta m_1 / m_1$} \\
 \hline 
 & & \multicolumn{3}{c|}{$5\%$ quantile} &
\multicolumn{3}{c|}{Median} & \multicolumn{3}{c|}{$95\%$
quantile} \\
 \hline 
 & & RWF & SWF & FWF & RWF & SWF & FWF & RWF & SWF & FWF \\
 \hline
 $3 \cdot 10^5$ & $10^5$ & $1.85 \cdot 10^{-4}$ & $1.69 \cdot 10^{-4}$ & $1.44
\cdot 10^{-4}$ & $8.06 \cdot 10^{-4}$ & $6.71 \cdot 10^{-4}$ & $4.76 \cdot
10^{-4}$ & $8.17 \cdot 10^{-3}$ & $2.54 \cdot 10^{-3}$ & $1.58 \cdot 10^{-3}$ \\
 \hline
 $10^6$ & $10^5$ &  $2.44 \cdot 10^{-4}$ & $1.99 \cdot 10^{-4}$ & $1.63 \cdot
10^{-4}$ & $7.13 \cdot 10^{-4}$ & $5.36 \cdot 10^{-4}$ & $4.25 \cdot 10^{-4}$ &
$4.63 \cdot 10^{-3}$ & $2.82 \cdot 10^{-3}$ & $2.13 \cdot 10^{-3}$ \\
 \hline
 $10^6$ & $3 \cdot 10^5$ & $4.08 \cdot 10^{-4}$ & $3.59 \cdot 10^{-4}$ & $2.90
\cdot 10^{-4}$ & $1.36 \cdot 10^{-3}$ & $1.10 \cdot 10^{-3}$ & $8.01 \cdot
10^{-4}$ & $1.11 \cdot 10^{-2}$ & $3.95 \cdot 10^{-3}$ & $2.60 \cdot 10^{-3}$ \\
 \hline \hline
 $3 \cdot 10^5$ & $3 \cdot 10^5$ & $1.69 \cdot 10^{-4}$ & $1.62 \cdot 10^{-4}$ &
$1.19 \cdot 10^{-4}$ & $1.24 \cdot 10^{-3}$ & $1.17 \cdot 10^{-3}$ & $2.91 \cdot
10^{-4}$ & $1.14 \cdot 10^{-2}$ & $8.82 \cdot 10^{-3}$ & $6.52 \cdot 10^{-4}$ \\
 \hline
 $10^6$ & $10^6$ & $3.59 \cdot 10^{-4}$ & $3.53 \cdot 10^{-4}$ & $2.53 \cdot
10^{-4}$ & $2.48 \cdot 10^{-3}$ & $2.42 \cdot 10^{-3}$ & $7.16 \cdot 10^{-4}$ &
$2.98 \cdot 10^{-2}$ & $2.29 \cdot 10^{-2}$ & $1.57 \cdot 10^{-3}$ \\
 \hline \hline
 $10^7$ & $10^6$ & $1.21 \cdot 10^{-3}$ & $7.93 \cdot 10^{-4}$ & $4.46 \cdot
10^{-4}$ & $4.34 \cdot 10^{-3}$ & $2.49 \cdot 10^{-3}$ & $1.37 \cdot 10^{-3}$ &
$4.21 \cdot 10^{-2}$ & $1.12 \cdot 10^{-2}$ & $5.09 \cdot 10^{-3}$ \\
 \hline
 $10^7$ & $3 \cdot 10^6$ & $3.39 \cdot 10^{-3}$ & $1.89 \cdot 10^{-3}$ & $8.23
\cdot 10^{-4}$ & $1.59 \cdot 10^{-2}$ & $4.26 \cdot 10^{-3}$ & $1.79 \cdot
10^{-3}$ & $0.140$ & $1.18 \cdot 10^{-2}$ & $5.86 \cdot 10^{-3}$ \\
 \hline
 $10^7$ & $10^7$ & $2.20 \cdot 10^{-2}$ & $1.21 \cdot 10^{-2}$ & $2.04 \cdot
10^{-3}$ & $0.213$ & $8.97 \cdot 10^{-2}$ & $5.79 \cdot 10^{-3}$ & $1.47$ &
$0.825$ & $1.57 \cdot 10^{-2}$ \\
 \hline
 $3 \cdot 10^7$ & $10^7$ & $0.377$ & $8.64 \cdot 10^{-3}$ & $4.35 \cdot 10^{-3}$
& $1.01$ & $1.99 \cdot 10^{-2}$ & $9.48 \cdot 10^{-3}$ & $3.23$ & $5.35 \cdot
10^{-2}$ & $2.43 \cdot 10^{-2}$ \\
 \hline
 $3 \cdot 10^7$ & $3 \cdot 10^7$ & $3.74$ & $0.525$ & $5.55 \cdot 10^{-2}$ &
$23.1$ & $2.26$ & $0.120$ & $115$ & $9.05$ & $0.386$ \\
 \hline
 $10^8$ & $10^7$ & $\infty$ & $9.67 \cdot 10^{-2}$ & $9.45 \cdot 10^{-2}$ &
$\infty$ & $0.276$ & $0.246$ & $\infty$ & $1.27$ & $1.00$ \\
 \hline
 $10^8$ & $3 \cdot 10^7$ & $\infty$ & $0.896$ & $0.963$ & $\infty$ & $2.57$ &
$2.86$ & $\infty$ & $53.6$ & $59.7$ \\
 \hline
\end{tabular}
\caption{Median, $5\%$ and $95\%$ quantiles of the estimated measurement errors
on $m_1$ for different
sets of binaries
 located at redshift $z=1$, with low unequal-mass binaries on top, low
equal-mass binaries in the middle, and high-mass binaries at the
bottom.\label{tab:m1}}
\end{table*}

\begin{table*}[!ht]
\begin{tabular}{|c|c|c|c|c|c|c|c|c|c|c|}
 \hline
 $m_1 [M_\odot]$ & $m_2 [M_\odot]$ & \multicolumn{9}{c|}{$\Delta m_2 / m_2$} \\
 \hline 
 & & \multicolumn{3}{c|}{$5\%$ quantile} &
\multicolumn{3}{c|}{Median} & \multicolumn{3}{c|}{$95\%$
quantile} \\
 \hline 
 & & RWF & SWF & FWF & RWF & SWF & FWF & RWF & SWF & FWF \\
 \hline
 $3 \cdot 10^5$ & $10^5$ & $1.50 \cdot 10^{-4}$ & $1.37 \cdot 10^{-4}$ & $1.18
\cdot 10^{-4}$ & $6.54 \cdot 10^{-4}$ & $5.44 \cdot 10^{-4}$ & $3.87 \cdot
10^{-4}$ & $6.64 \cdot 10^{-3}$ & $2.06 \cdot 10^{-3}$ & $1.28 \cdot 10^{-3}$ \\
 \hline
 $10^6$ & $10^5$ & $1.76 \cdot 10^{-4}$ & $1.41 \cdot 10^{-4}$ & $1.18 \cdot
10^{-4}$ & $5.05 \cdot 10^{-4}$ & $3.78 \cdot 10^{-4}$ & $3.03 \cdot 10^{-4}$ &
$3.27 \cdot 10^{-3}$ & $2.00 \cdot 10^{-3}$ & $1.50 \cdot 10^{-3}$ \\
 \hline
 $10^6$ & $3 \cdot 10^5$ & $3.26 \cdot 10^{-4}$ & $2.89 \cdot 10^{-4}$ & $2.33
\cdot 10^{-4}$ & $1.08 \cdot 10^{-3}$ & $8.84 \cdot 10^{-4}$ & $6.41 \cdot
10^{-4}$ & $8.91 \cdot 10^{-3}$ & $3.15 \cdot 10^{-3}$ & $2.09 \cdot 10^{-3}$ \\
 \hline \hline
 $3 \cdot 10^5$ & $3 \cdot 10^5$ & $1.64 \cdot 10^{-4}$ & $1.59 \cdot 10^{-4}$ &
$1.21 \cdot 10^{-4}$ & $1.25 \cdot 10^{-3}$ & $1.17 \cdot 10^{-3}$ & $2.90 \cdot
10^{-4}$ & $1.15 \cdot 10^{-2}$ & $8.75 \cdot 10^{-3}$ & $6.55 \cdot 10^{-4}$ \\
 \hline
 $10^6$ & $10^6$ & $3.53 \cdot 10^{-4}$ & $3.48 \cdot 10^{-4}$ & $2.56 \cdot
10^{-4}$ & $2.52 \cdot 10^{-3}$ & $2.41 \cdot 10^{-3}$ & $7.13 \cdot 10^{-4}$ &
$2.97 \cdot 10^{-2}$ & $2.30 \cdot 10^{-2}$ & $1.57 \cdot 10^{-3}$ \\
 \hline \hline
 $10^7$ & $10^6$ & $1.19 \cdot 10^{-3}$ & $7.20 \cdot 10^{-4}$ & $4.25 \cdot
10^{-4}$ & $3.48 \cdot 10^{-3}$ & $1.84 \cdot 10^{-3}$ & $1.04 \cdot 10^{-3}$ &
$2.94 \cdot 10^{-2}$ & $7.89 \cdot 10^{-3}$ & $3.59 \cdot 10^{-3}$ \\
 \hline
 $10^7$ & $3 \cdot 10^6$ & $3.20 \cdot 10^{-3}$ & $1.66 \cdot 10^{-3}$ & $7.45
\cdot 10^{-4}$ & $1.33 \cdot 10^{-2}$ & $3.56 \cdot 10^{-3}$ & $1.54 \cdot
10^{-3}$ & $0.110$ & $9.44 \cdot 10^{-3}$ & $4.74 \cdot 10^{-3}$ \\
 \hline
 $10^7$ & $10^7$ & $2.20 \cdot 10^{-2}$ & $1.20 \cdot 10^{-2}$ & $2.05 \cdot
10^{-3}$ & $0.208$ & $9.15 \cdot 10^{-2}$ & $5.79 \cdot 10^{-3}$ & $1.49$ &
$0.820$ & $1.57 \cdot 10^{-2}$ \\
 \hline
 $3 \cdot 10^7$ & $10^7$ & $0.412$ & $1.06 \cdot 10^{-2}$ & $5.78 \cdot 10^{-3}$
& $1.57$ & $2.34 \cdot 10^{-2}$ & $1.30 \cdot 10^{-2}$ & $4.91$ & $5.54 \cdot
10^{-2}$ & $3.27 \cdot 10^{-2}$ \\
 \hline
 $3 \cdot 10^7$ & $3 \cdot 10^7$ & $3.57$ & $0.557$ & $5.54 \cdot 10^{-2}$ &
$23.0$ & $2.21$ & $0.120$ & $108$ & $8.89$ & $0.386$ \\
 \hline
 $10^8$ & $10^7$ & $\infty$ & $0.264$ & $0.336$ & $\infty$ & $0.867$ & $1.05$ &
$\infty$ & $3.23$ & $3.95$ \\
 \hline
 $10^8$ & $3 \cdot 10^7$ & $\infty$ & $3.19$ & $3.56$ & $\infty$ & $9.75$ &
$10.3$ & $\infty$ & $145$ & $160$ \\
 \hline
\end{tabular}
\caption{Median, $5\%$ and $95\%$ quantiles of the estimated measurement errors
on $m_2$ for different
sets of binaries
 located at redshift $z=1$, with low unequal-mass binaries on top, low
equal-mass binaries in the middle, and high-mass binaries at the
bottom.\label{tab:m2}}
\end{table*}

\begin{table*}[!ht]
\begin{tabular}{|c|c|c|c|c|c|c|c|c|c|c|}
 \hline
 $m_1 [M_\odot]$ & $m_2 [M_\odot]$ & \multicolumn{9}{c|}{$\Delta \chi_1$} \\
 \hline 
 & & \multicolumn{3}{c|}{$5\%$ quantile} &
\multicolumn{3}{c|}{Median} & \multicolumn{3}{c|}{$95\%$
quantile} \\
 \hline 
 & & RWF & SWF & FWF & RWF & SWF & FWF & RWF & SWF & FWF \\
 \hline
 $3 \cdot 10^5$ & $10^5$ & $4.95 \cdot 10^{-4}$ & $4.11 \cdot 10^{-4}$ & $2.94
\cdot 10^{-4}$ & $1.47 \cdot 10^{-3}$ & $1.17 \cdot 10^{-3}$ & $8.12 \cdot
10^{-4}$ & $9.63 \cdot 10^{-3}$ & $4.83 \cdot 10^{-3}$ & $3.32 \cdot 10^{-3}$ \\
 \hline
 $10^6$ & $10^5$ & $3.81 \cdot 10^{-4}$ & $2.72 \cdot 10^{-4}$ & $1.97 \cdot
10^{-4}$ & $8.98 \cdot 10^{-4}$ & $6.11 \cdot 10^{-4}$ & $4.38 \cdot 10^{-4}$ &
$2.72 \cdot 10^{-3}$ & $1.78 \cdot 10^{-3}$ & $1.39 \cdot 10^{-3}$ \\
 \hline
 $10^6$ & $3 \cdot 10^5$ & $8.68 \cdot 10^{-4}$ & $7.30 \cdot 10^{-4}$ & $4.97
\cdot 10^{-4}$ & $2.16 \cdot 10^{-3}$ & $1.69 \cdot 10^{-3}$ & $1.20 \cdot
10^{-3}$ & $1.05 \cdot 10^{-2}$ & $5.57 \cdot 10^{-3}$ & $3.93 \cdot 10^{-3}$ \\
 \hline \hline
 $3 \cdot 10^5$ & $3 \cdot 10^5$ & $9.96 \cdot 10^{-4}$ & $9.62 \cdot 10^{-4}$ &
$7.46 \cdot 10^{-4}$ & $6.81 \cdot 10^{-3}$ & $6.45 \cdot 10^{-3}$ & $3.93 \cdot
10^{-3}$ & $8.34 \cdot 10^{-2}$ & $7.08 \cdot 10^{-2}$ & $3.75 \cdot 10^{-2}$ \\
 \hline
 $10^6$ & $10^6$ & $1.53 \cdot 10^{-3}$ & $1.50 \cdot 10^{-3}$ & $1.17 \cdot
10^{-3}$ & $1.31 \cdot 10^{-2}$ & $1.27 \cdot 10^{-2}$ & $6.82 \cdot 10^{-3}$ &
$0.218$ & $0.189$ & $7.51 \cdot 10^{-2}$ \\
 \hline \hline
 $10^7$ & $10^6$ & $1.37 \cdot 10^{-3}$ & $8.44 \cdot 10^{-4}$ & $4.58 \cdot
10^{-4}$ & $3.68 \cdot 10^{-3}$ & $1.90 \cdot 10^{-3}$ & $1.07 \cdot 10^{-3}$ &
$1.64 \cdot 10^{-2}$ & $6.33 \cdot 10^{-3}$ & $3.45 \cdot 10^{-3}$ \\
 \hline
 $10^7$ & $3 \cdot 10^6$ & $3.87 \cdot 10^{-3}$ & $2.41 \cdot 10^{-3}$ & $1.21
\cdot 10^{-3}$ & $1.47 \cdot 10^{-2}$ & $5.93 \cdot 10^{-3}$ & $3.03 \cdot
10^{-3}$ & $0.118$ & $2.15 \cdot 10^{-2}$ & $1.24 \cdot 10^{-2}$ \\
 \hline
 $10^7$ & $10^7$ & $0.108$ & $5.00 \cdot 10^{-2}$ & $2.04 \cdot 10^{-2}$ &
$1.20$ & $0.488$ & $0.136$ & $9.77$ & $5.45$ & $1.17$ \\
 \hline
 $3 \cdot 10^7$ & $10^7$ & $0.438$ & $1.93 \cdot 10^{-2}$ & $9.36 \cdot 10^{-3}$
& $1.87$ & $6.19 \cdot 10^{-2}$ & $3.31 \cdot 10^{-2}$ & $7.24$ & $0.269$ &
$0.147$ \\
 \hline
 $3 \cdot 10^7$ & $3 \cdot 10^7$ & $16.8$ & $2.04$ & $1.31$ & $83.6$ & $11.9$ &
$5.35$ & $499$ & $61.4$ & $26.3$ \\
 \hline
 $10^8$ & $10^7$ & $\infty$ & $0.256$ & $0.316$ & $\infty$ & $1.10$ & $1.35$ &
$\infty$ & $4.65$ & $5.62$ \\
 \hline
 $10^8$ & $3 \cdot 10^7$ & $\infty$ & $3.47$ & $4.19$ & $\infty$ & $15.1$ &
$16.3$ & $\infty$ & $172$ & $185$ \\
 \hline
\end{tabular}
\caption{Median, $5\%$ and $95\%$ quantiles of the estimated measurement errors
on $\chi_1$ for different
sets of binaries
 located at redshift $z=1$, with low unequal-mass binaries on top, low
equal-mass binaries in the middle, and high-mass binaries at the
bottom.\label{tab:ch1}}
\end{table*}

\begin{table*}[!ht]
\begin{tabular}{|c|c|c|c|c|c|c|c|c|c|c|}
 \hline
 $m_1 [M_\odot]$ & $m_2 [M_\odot]$ & \multicolumn{9}{c|}{$\Delta \chi_2$} \\
 \hline 
 & & \multicolumn{3}{c|}{$5\%$ quantile} &
\multicolumn{3}{c|}{Median} & \multicolumn{3}{c|}{$95\%$
quantile} \\
 \hline 
 & & RWF & SWF & FWF & RWF & SWF & FWF & RWF & SWF & FWF \\
 \hline
 $3 \cdot 10^5$ & $10^5$ & $8.05 \cdot 10^{-4}$ & $7.19 \cdot 10^{-4}$ & $5.40
\cdot 10^{-4}$ & $3.23 \cdot 10^{-3}$ & $2.70 \cdot 10^{-3}$ & $1.91 \cdot
10^{-3}$ & $2.30 \cdot 10^{-2}$ & $1.34 \cdot 10^{-2}$ & $9.81 \cdot 10^{-3}$ \\
 \hline
 $10^6$ & $10^5$ & $1.71 \cdot 10^{-3}$ & $1.09 \cdot 10^{-3}$ & $8.34 \cdot
10^{-4}$ & $5.24 \cdot 10^{-3}$ & $3.61 \cdot 10^{-3}$ & $2.78 \cdot 10^{-3}$ &
$3.20 \cdot 10^{-2}$ & $2.00 \cdot 10^{-2}$ & $1.45 \cdot 10^{-2}$ \\
 \hline
 $10^6$ & $3 \cdot 10^5$ & $1.41 \cdot 10^{-3}$ & $1.22 \cdot 10^{-3}$ & $9.01
\cdot 10^{-4}$ & $4.72 \cdot 10^{-3}$ & $3.91 \cdot 10^{-3}$ & $2.78 \cdot
10^{-3}$ & $2.65 \cdot 10^{-2}$ & $1.63 \cdot 10^{-2}$ & $1.20 \cdot 10^{-2}$ \\
 \hline \hline
 $3 \cdot 10^5$ & $3 \cdot 10^5$ & $1.02 \cdot 10^{-3}$ & $9.58 \cdot 10^{-4}$ &
$7.83 \cdot 10^{-4}$ & $6.66 \cdot 10^{-3}$ & $6.34 \cdot 10^{-3}$ & $3.95 \cdot
10^{-3}$ & $7.84 \cdot 10^{-2}$ & $6.33 \cdot 10^{-2}$ & $3.68 \cdot 10^{-2}$ \\
 \hline
 $10^6$ & $10^6$ & $1.67 \cdot 10^{-3}$ & $1.66 \cdot 10^{-3}$ & $1.22 \cdot
10^{-3}$ & $1.27 \cdot 10^{-2}$ & $1.25 \cdot 10^{-2}$ & $6.76 \cdot 10^{-3}$ &
$0.217$ & $0.186$ & $7.07 \cdot 10^{-2}$ \\
 \hline \hline
 $10^7$ & $10^6$ & $6.33 \cdot 10^{-3}$ & $3.68 \cdot 10^{-3}$ & $1.98 \cdot
10^{-3}$ & $3.31 \cdot 10^{-2}$ & $1.63 \cdot 10^{-2}$ & $9.66 \cdot 10^{-3}$ &
$0.187$ & $7.57 \cdot 10^{-2}$ & $4.01 \cdot 10^{-2}$ \\
 \hline
 $10^7$ & $3 \cdot 10^6$ & $7.35 \cdot 10^{-3}$ & $4.64 \cdot 10^{-3}$ & $2.42
\cdot 10^{-3}$ & $3.32 \cdot 10^{-2}$ & $1.72 \cdot 10^{-2}$ & $9.46 \cdot
10^{-3}$ & $0.193$ & $6.87 \cdot 10^{-2}$ & $4.13 \cdot 10^{-2}$ \\
 \hline
 $10^7$ & $10^7$ & $0.111$ & $4.66 \cdot 10^{-2}$ & $1.94 \cdot 10^{-2}$ &
$1.22$ & $0.511$ & $0.130$ & $9.94$ & $5.60$ & $1.11$ \\
 \hline
 $3 \cdot 10^7$ & $10^7$ & $0.594$ & $4.02 \cdot 10^{-2}$ & $2.07 \cdot 10^{-2}$
& $4.44$ & $0.205$ & $0.104$ & $26.8$ & $0.960$ & $0.496$ \\
 \hline
 $3 \cdot 10^7$ & $3 \cdot 10^7$ & $16.1$ & $2.20$ & $1.34$ & $83.0$ & $11.4$ &
$5.43$ & $515$ & $59.5$ & $26.5$ \\
 \hline
 $10^8$ & $10^7$ & $\infty$ & $1.07$ & $1.38$ & $\infty$ & $11.7$ & $13.7$ &
$\infty$ & $52.8$ & $62.6$ \\
 \hline
 $10^8$ & $3 \cdot 10^7$ & $\infty$ & $8.31$ & $9.25$ & $\infty$ & $48.0$ &
$50.4$ & $\infty$ & $657$ & $700$ \\
 \hline
\end{tabular}
\caption{Median, $5\%$ and $95\%$ quantiles of the estimated measurement errors
on $\chi_2$ for different
sets of binaries
 located at redshift $z=1$, with low unequal-mass binaries on top, low
equal-mass binaries in the middle, and high-mass binaries at the
bottom.\label{tab:ch2}}
\end{table*}

\begin{table*}[!ht]
\begin{tabular}{|c|c|c|c|c|c|c|c|c|c|c|}
 \hline
 $m_1 [M_\odot]$ & $m_2 [M_\odot]$ & \multicolumn{9}{c|}{$2a~[']$} \\
 \hline 
 & & \multicolumn{3}{c|}{$5\%$ quantile} &
\multicolumn{3}{c|}{Median} & \multicolumn{3}{c|}{$95\%$
quantile} \\
 \hline 
 & & RWF & SWF & FWF & RWF & SWF & FWF & RWF & SWF & FWF \\
 \hline
 $3 \cdot 10^5$ & $10^5$ & $5.09$ & $4.96$ & $3.18$ & $20.2$ & $18.5$ & $12.7$ &
$92.2$ & $85.8$ & $67.2$ \\
 \hline
 $10^6$ & $10^5$ & $8.67$ & $8.13$ & $7.14$ & $34.1$ & $28.2$ & $20.1$ & $124$ &
$100$ & $82.5$ \\
 \hline
 $10^6$ & $3 \cdot 10^5$ & $8.95$ & $8.61$ & $6.28$ & $31.4$ & $28.0$ & $19.8$ &
$124$ & $110$ & $85.0$ \\
 \hline \hline
 $3 \cdot 10^5$ & $3 \cdot 10^5$ & $6.00$ & $5.81$ & $3.73$ & $26.6$ & $25.4$ &
$18.6$ & $113$ & $109$ & $97.7$ \\
 \hline
 $10^6$ & $10^6$ & $8.45$ & $8.40$ & $5.42$ & $38.5$ & $37.6$ & $26.3$ & $158$ &
$154$ & $129$ \\
 \hline \hline
 $10^7$ & $10^6$ & $19.2$ & $15.5$ & $8.34$ & $64.2$ & $48.4$ & $23.2$ & $316$ &
$199$ & $113$ \\
 \hline
 $10^7$ & $3 \cdot 10^6$ & $19.5$ & $17.4$ & $7.31$ & $84.3$ & $65.2$ & $31.0$ &
$461$ & $283$ & $158$ \\
 \hline
 $10^7$ & $10^7$ & $32.7$ & $27.4$ & $11.6$ & $202$ & $155$ & $77.7$ & $1360$ &
$818$ & $496$ \\
 \hline
 $3 \cdot 10^7$ & $10^7$ & $169$ & $68.5$ & $24.4$ & $1500$ & $319$ & $133$ &
$16600$ & $1550$ & $738$ \\
 \hline
 $3 \cdot 10^7$ & $3 \cdot 10^7$ & $7910$ & $537$ & $363$ & $188000$ & $2590$ &
$2570$ & $2890000$ & $14700$ & $16400$ \\
 \hline
 $10^8$ & $10^7$ & $\infty$ & $998$ & $1300$ & $\infty$ & $3380$ &
$4400$ & $\infty$ & $18200$ & $23800$ \\
 \hline
 $10^8$ & $3 \cdot 10^7$ & $\infty$ & $5900$ & $6510$ & $\infty$ & $26300$ &
$31000$ & $\infty$ & $237000$ & $279000$ \\
 \hline
\end{tabular}
\caption{Median, $5\%$ and $95\%$ quantiles of the estimated measurement errors
on the major axis of the localization ellipse in the sky for different
sets of binaries
 located at redshift $z=1$, with low unequal-mass binaries on top, low
equal-mass binaries in the middle, and high-mass binaries at the
bottom.\label{tab:2a}}
\end{table*}

\begin{table*}[!ht]
\begin{tabular}{|c|c|c|c|c|c|c|c|c|c|c|}
 \hline
 $m_1 [M_\odot]$ & $m_2 [M_\odot]$ & \multicolumn{9}{c|}{$2b~[']$} \\
 \hline 
 & & \multicolumn{3}{c|}{$5\%$ quantile} &
\multicolumn{3}{c|}{Median} & \multicolumn{3}{c|}{$95\%$
quantile} \\
 \hline 
 & & RWF & SWF & FWF & RWF & SWF & FWF & RWF & SWF & FWF \\
 \hline
 $3 \cdot 10^5$ & $10^5$ & $0.795$ & $0.778$ & $0.453$ & $3.76$ & $3.49$ &
$2.10$ & $13.8$ & $12.1$ & $7.40$ \\
 \hline
 $10^6$ & $10^5$ & $2.17$ & $1.69$ & $0.985$ & $10.2$ & $7.61$ & $4.50$ & $23.6$
& $15.5$ & $10.2$ \\
 \hline
 $10^6$ & $3 \cdot 10^5$ & $1.83$ & $1.63$ & $0.984$ & $8.63$ & $7.70$ & $4.47$
& $24.4$ & $19.2$ & $12.5$ \\
 \hline \hline
 $3 \cdot 10^5$ & $3 \cdot 10^5$ & $1.00$ & $1.00$ & $0.575$ & $5.52$ & $5.29$ &
$3.17$ & $20.3$ & $18.6$ & $13.8$ \\
 \hline
 $10^6$ & $10^6$ & $1.62$ & $1.61$ & $0.948$ & $9.11$ & $9.04$ & $5.26$ & $31.9$
& $29.9$ & $19.4$ \\
 \hline \hline
 $10^7$ & $10^6$ & $3.40$ & $2.81$ & $1.17$ & $15.7$ & $12.5$ & $5.20$ & $41.3$
& $27.2$ & $12.3$ \\
 \hline
 $10^7$ & $3 \cdot 10^6$ & $3.04$ & $2.65$ & $1.11$ & $13.8$ & $11.9$ & $4.87$ &
$54.3$ & $37.2$ & $17.3$ \\
 \hline
 $10^7$ & $10^7$ & $5.77$ & $4.64$ & $1.93$ & $24.8$ & $21.6$ & $9.49$ & $125$ &
$90.8$ & $48.3$ \\
 \hline
 $3 \cdot 10^7$ & $10^7$ & $36.4$ & $11.1$ & $4.00$ & $164$ & $47.9$ & $17.5$ &
$896$ & $157$ & $69.9$ \\
 \hline
 $3 \cdot 10^7$ & $3 \cdot 10^7$ & $727$ & $108$ & $57.2$ & $5740$ & $306$ &
$230$ & $85000$ & $1090$ & $1350$ \\
 \hline
 $10^8$ & $10^7$ & $\infty$ & $201$ & $251$ & $\infty$ & $724$ & $930$ &
$\infty$ & $2610$ & $3600$ \\
 \hline
 $10^8$ & $3 \cdot 10^7$ & $\infty$ & $1140$ & $1340$ & $\infty$ & $3890$ &
$4630$ & $\infty$ & $44400$ & $52500$ \\
 \hline
\end{tabular}
\caption{Median, $5\%$ and $95\%$ quantiles of the estimated measurement errors
on the minor axis of the localization ellipse in the sky for different
sets of binaries
 located at redshift $z=1$, with low unequal-mass binaries on top, low
equal-mass binaries in the middle, and high-mass binaries at the
bottom.\label{tab:2b}}
\end{table*}

\begin{table*}[!ht]
\begin{tabular}{|c|c|c|c|c|c|c|c|c|c|c|}
 \hline
 $m_1 [M_\odot]$ & $m_2 [M_\odot]$ & \multicolumn{9}{c|}{$\Delta d_L / d_L$} \\
 \hline 
 & & \multicolumn{3}{c|}{$5\%$ quantile} &
\multicolumn{3}{c|}{Median} & \multicolumn{3}{c|}{$95\%$
quantile} \\
 \hline 
 & & RWF & SWF & FWF & RWF & SWF & FWF & RWF & SWF & FWF \\
 \hline
 $3 \cdot 10^5$ & $10^5$ & $8.67 \cdot 10^{-4}$ & $8.24 \cdot 10^{-4}$ & $5.76
\cdot 10^{-4}$ & $2.94 \cdot 10^{-3}$ & $2.43 \cdot 10^{-3}$ & $1.75 \cdot
10^{-3}$ & $1.28 \cdot 10^{-2}$ & $1.01 \cdot 10^{-2}$ & $8.43 \cdot 10^{-3}$ \\
 \hline
 $10^6$ & $10^5$ & $2.08 \cdot 10^{-3}$ & $1.44 \cdot 10^{-3}$ & $1.03 \cdot
10^{-3}$ & $4.90 \cdot 10^{-3}$ & $3.51 \cdot 10^{-3}$ & $2.44 \cdot 10^{-3}$ &
$1.49 \cdot 10^{-2}$ & $8.74 \cdot 10^{-3}$ & $7.04 \cdot 10^{-3}$ \\
 \hline
 $10^6$ & $3 \cdot 10^5$ & $1.75 \cdot 10^{-3}$ & $1.43 \cdot 10^{-3}$ & $1.11
\cdot 10^{-3}$ & $4.62 \cdot 10^{-3}$ & $3.74 \cdot 10^{-3}$ & $2.70 \cdot
10^{-3}$ & $1.97 \cdot 10^{-2}$ & $1.39 \cdot 10^{-2}$ & $1.12 \cdot 10^{-2}$ \\
 \hline \hline
 $3 \cdot 10^5$ & $3 \cdot 10^5$ & $1.13 \cdot 10^{-3}$ & $1.08 \cdot 10^{-3}$ &
$7.20 \cdot 10^{-4}$ & $4.41 \cdot 10^{-3}$ & $3.80 \cdot 10^{-3}$ & $2.88 \cdot
10^{-3}$ & $1.89 \cdot 10^{-2}$ & $1.48 \cdot 10^{-2}$ & $1.27 \cdot 10^{-2}$ \\
 \hline
 $10^6$ & $10^6$ & $1.85 \cdot 10^{-3}$ & $1.77 \cdot 10^{-3}$ & $1.12 \cdot
10^{-3}$ & $6.88 \cdot 10^{-3}$ & $6.20 \cdot 10^{-3}$ & $4.36 \cdot 10^{-3}$ &
$2.65 \cdot 10^{-2}$ & $2.02 \cdot 10^{-2}$ & $1.71 \cdot 10^{-2}$ \\
 \hline \hline
 $10^7$ & $10^6$ & $3.14 \cdot 10^{-3}$ & $2.32 \cdot 10^{-3}$ & $1.48 \cdot
10^{-3}$ & $8.53 \cdot 10^{-3}$ & $6.25 \cdot 10^{-3}$ & $3.45 \cdot 10^{-3}$ &
$4.14 \cdot 10^{-2}$ & $1.94 \cdot 10^{-2}$ & $1.26 \cdot 10^{-2}$ \\
 \hline
 $10^7$ & $3 \cdot 10^6$ & $3.84 \cdot 10^{-3}$ & $3.11 \cdot 10^{-3}$ & $2.18
\cdot 10^{-3}$ & $1.09 \cdot 10^{-2}$ & $8.03 \cdot 10^{-3}$ & $4.69 \cdot
10^{-3}$ & $7.26 \cdot 10^{-2}$ & $3.43 \cdot 10^{-2}$ & $1.99 \cdot 10^{-2}$ \\
 \hline
 $10^7$ & $10^7$ & $1.45 \cdot 10^{-2}$ & $8.99 \cdot 10^{-3}$ & $6.78 \cdot
10^{-3}$ & $6.05 \cdot 10^{-2}$ & $2.76 \cdot 10^{-2}$ & $2.03 \cdot 10^{-2}$ &
$0.277$ & $0.113$ & $7.82 \cdot 10^{-2}$ \\
 \hline
 $3 \cdot 10^7$ & $10^7$ & $0.212$ & $1.94 \cdot 10^{-2}$ & $1.68 \cdot 10^{-2}$
& $0.801$ & $4.63 \cdot 10^{-2}$ & $3.52 \cdot 10^{-2}$ & $3.44$ & $0.186$ &
$0.101$ \\
 \hline
 $3 \cdot 10^7$ & $3 \cdot 10^7$ & $3.52$ & $0.188$ & $0.213$ & $31.7$ & $0.614$
& $0.523$ & $462$ & $2.33$ & $2.23$ \\
 \hline
 $10^8$ & $10^7$ & $\infty$ & $0.257$ & $0.424$ & $\infty$ & $0.698$ & $1.11$ &
$\infty$ & $2.53$ & $4.00$ \\
 \hline
 $10^8$ & $3 \cdot 10^7$ & $\infty$ & $1.76$ & $3.17$ & $\infty$ & $5.53$ &
$9.16$ & $\infty$ & $69.7$ & $111$ \\
 \hline
\end{tabular}
\caption{Median, $5\%$ and $95\%$ quantiles of the estimated measurement errors
on $d_L$ for different
sets of binaries
 located at redshift $z=1$, with low unequal-mass binaries on top, low
equal-mass binaries in the middle, and high-mass binaries at the
bottom.\label{tab:dl}}
\end{table*}

We found that the binaries can roughly be separated into three classes: low
unequal-mass binaries, low equal-mass binaries ($M \lesssim 10^7 M_\odot$), and
high-mass binaries ($M \gtrsim 10^7 M_\odot$). We discuss below these three
distinct cases, and plot the estimated error distributions for a representative
sample of each one of the three classes on the parameters $\Delta m_1/m_1$, 
$\Delta
\chi_1$, $2a$, and
$\Delta d_L/d_L$. The distributions for $\Delta m_2/m_2$ are similar to those
for $\Delta m_1/m_1$, those for $\Delta\chi_2$ to those for
$\Delta\chi_1$, and those for $2b$ to those for $2a$.

In general, for lower-mass binaries and independently on the mass ratio, we
find that the errors expected for extrinsic parameters using the FWF are $\sim
1.5$ times the ones expected for the RWF. This factor is $\sim 1.2$ comparing
the SWF to the RWF. This is changed when considering higher-mass
binaries, because the second harmonic, the only one present in the RWF, spends
very few cycles inside the LISA band.

To discuss the mass limit above which no information can be extracted from a
system anymore, we
present at the end the proportion of systems for which the
individual masses
and the luminosity distance can be measured with $50\%$ and $25\%$ accuracy,
for
all samples. We also plot for different mass ratios the maximum redshift at
which information can be extracted from a binary system, as a function of $m_1$.

We present then for each waveform how far the measurement of supermassive black
hole mergers could help determining the Hubble diagram. To do so, we compute up
to what redshift half of the systems can be localized inside the field of view
of Hubble and/or XMM-Newton (see e.g.~\cite{csmq}) which we take to be $30'$
wide, with an error on $d_L$ smaller than $10\%$.

\subsection{Low unequal-mass binaries}

\begin{figure}[!ht]
 \includegraphics[width=\columnwidth]{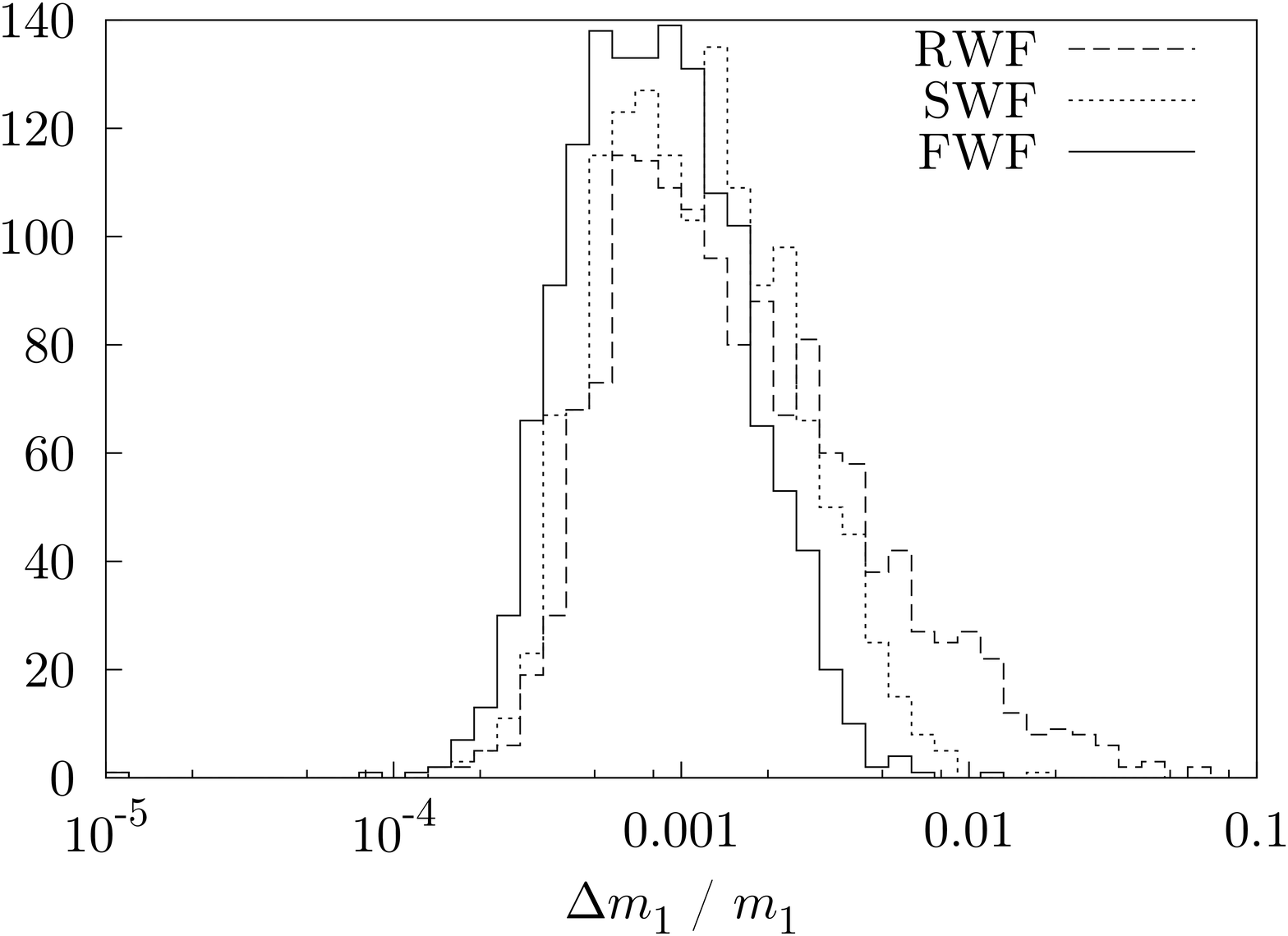}
  \caption{Estimated distribution of the measurement error on
 $m_1$ for a low
unequal-mass binary system with $m_1 = 10^6 M_\odot$ and $m_2 = 3
\cdot 10^5 M_\odot$. We expect to have errors as high as $8 \cdot 10^{-3}$ with
the RWF ($95\%$ percentile), whereas we do not expect errors higher than $1.5
\cdot 10^{-3}$ with the FWF.\label{fig:m1635}}
\end{figure}

\begin{figure}[!ht]
  \includegraphics[width=\columnwidth]{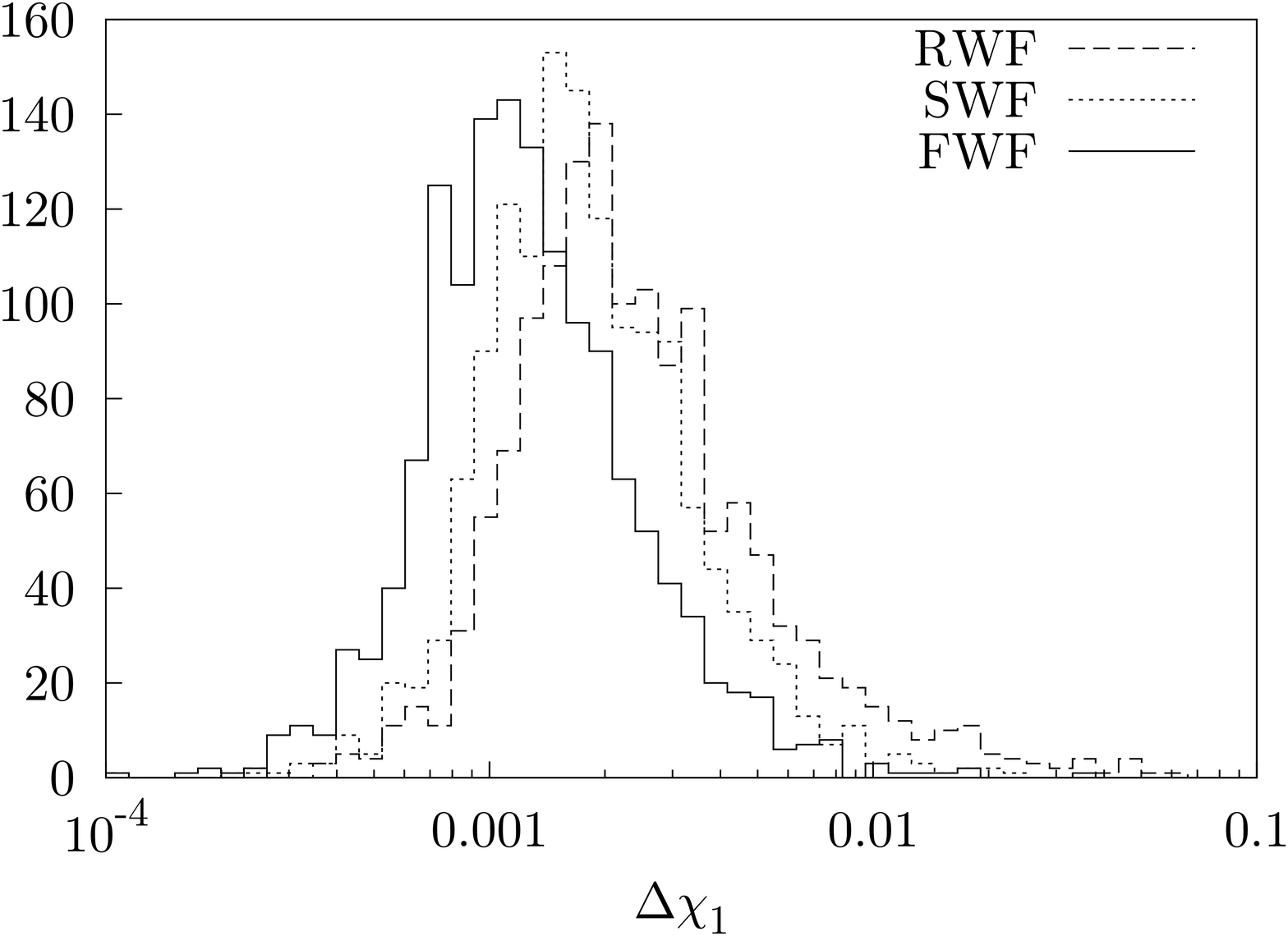}
  \caption{Estimated distribution of the measurement error on
 $\chi_1$ for a low
unequal-mass binary system with $m_1 = 10^6 M_\odot$ and $m_2 = 3
\cdot 10^5 M_\odot$. We expect the error to be $1.5 \upto 2$ times
lower using the FWF than using the RWF.\label{fig:ch1635}}
\end{figure}

\begin{figure}[!ht]
  \includegraphics[width=\columnwidth]{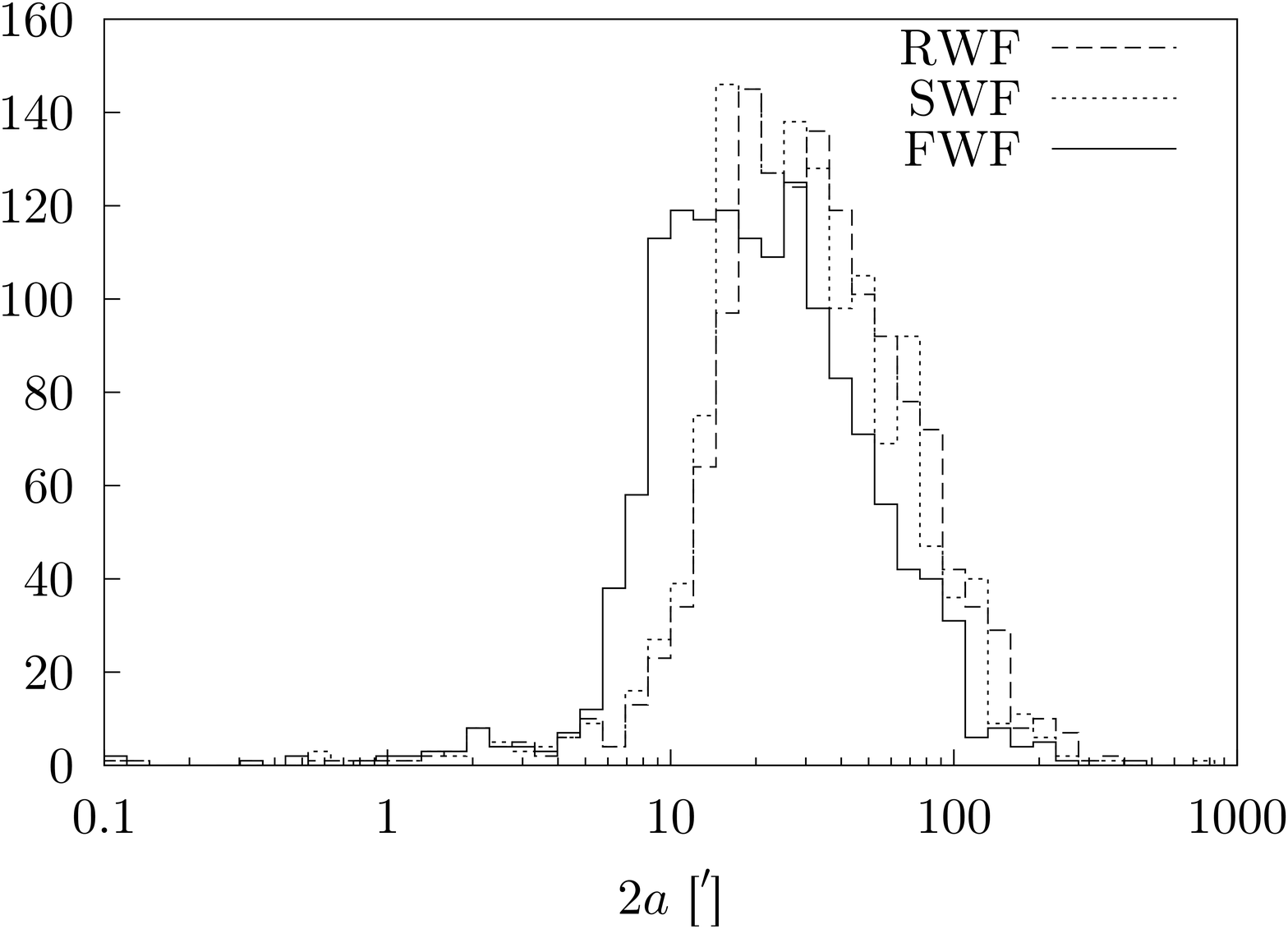}
  \caption{Estimated distribution of the major axis of the positioning error
ellipse for a low
unequal-mass binary system with $m_1 = 10^6 M_\odot$ and $m_2 = 3
\cdot 10^5 M_\odot$. We expect the error to be $\sim 1.5$ times
lower using the FWF than using the RWF.\label{fig:2a1635}}
\end{figure}

\begin{figure}[!ht]
  \includegraphics[width=\columnwidth]{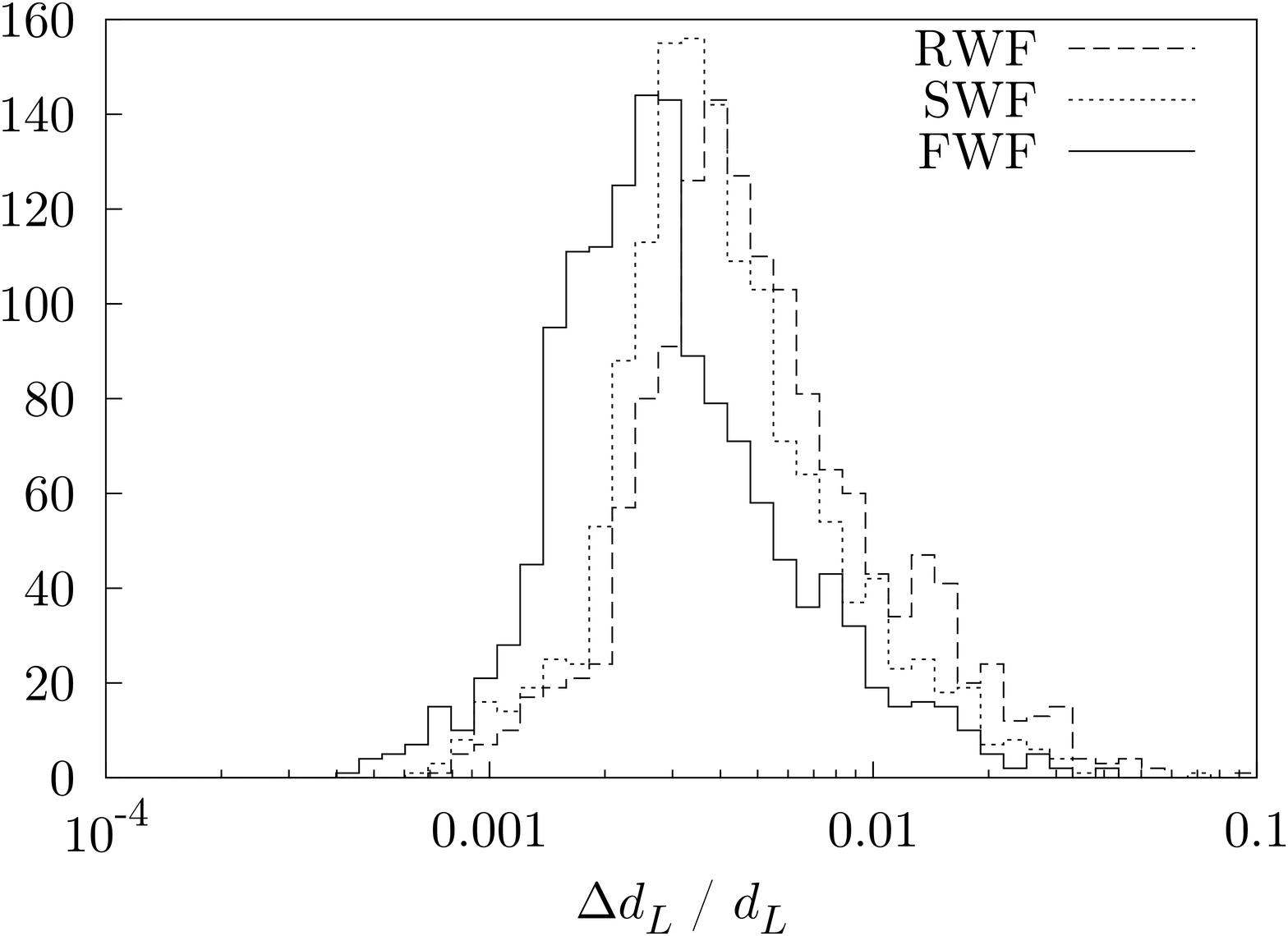}
  \caption{Estimated distribution of the measurement error on
 $d_L$ for a low
unequal-mass binary systems with $m_1 = 10^6 M_\odot$ and $m_2 = 3
\cdot 10^5 M_\odot$. We expect the error to be $\sim 1.5$ times
lower using the FWF than using the RWF.\label{fig:dl1635}}
\end{figure}

We put in this class all systems with total mass smaller than $10^7 M_\odot$,
and with a mass ratio of at least 1:3. We chose to present as a
representative sample systems with $m_1 = 10^6 M_\odot$ and $m_2 = 3 \cdot 10^5
M_\odot$. We plot the estimated distribution of the errors on $m_1$ in
Fig.~\ref{fig:m1635}, on $\chi_1$ in Fig.~\ref{fig:ch1635}, on the sky
positioning in Fig.~\ref{fig:2a1635}, and on $d_L$ in
Fig.~\ref{fig:dl1635}.

For these systems, the gain in accuracy obtained in the determination of all
interesting parameters with respect
to the RWF is typically a factor $\sim 1.5$ for the FWF and a factor $\sim 1.2$
for the SWF. However, when the mass ratio is close to 1:3,
the distribution of the errors on the individual masses for the RWF has a
relatively long tail of bad errors, which is
absent for the SWF and FWF.

The fact that including such extra structure as contained in the FWF
fails to provide much extra accuracy can allow including extra parameters in
the template. It has been recently suggested that the eccentricity of SMBH
binaries could be significant in the last stages of the inspiral~\cite{bpbms}.
Thus, inserting eccentricity parameters~\cite{yabw} could be important.
Furthermore, GW observations could help constraining alternative gravity
theories~\cite{bbw,arunwill,stavridiswill}.

\subsection{Low equal-mass binaries}

\begin{figure}[!ht]
  \includegraphics[width=\columnwidth]{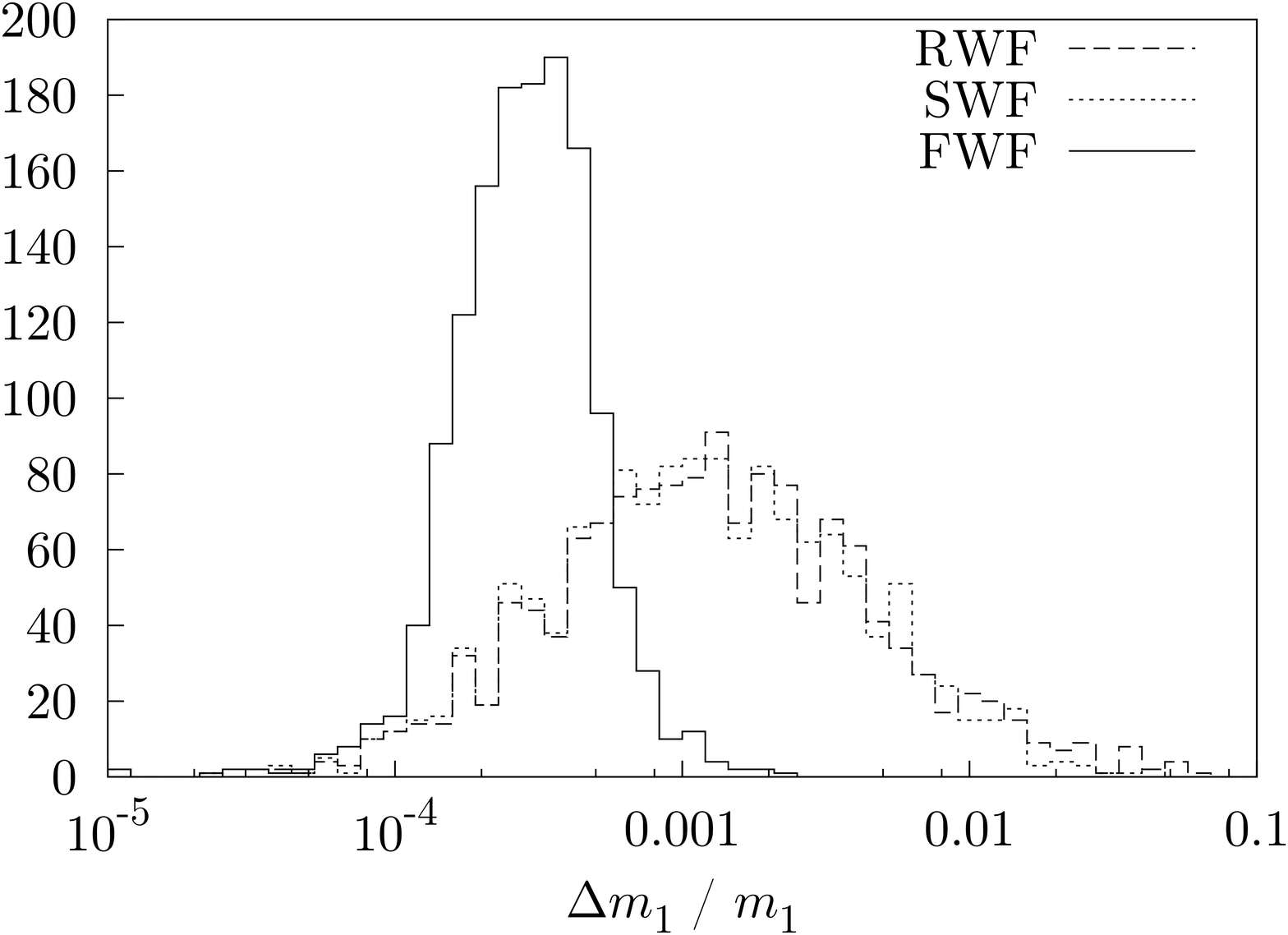}
  \caption{Estimated distribution of the measurement error on
 $m_1$ for a low equal-mass binary system with $m_1 = m_2 = 3 \cdot 10^5
M_\odot$. The FWF clearly gives better results than the two
other waveforms.\label{fig:m3535}}
\end{figure}

\begin{figure}[!ht]
  \includegraphics[width=\columnwidth]{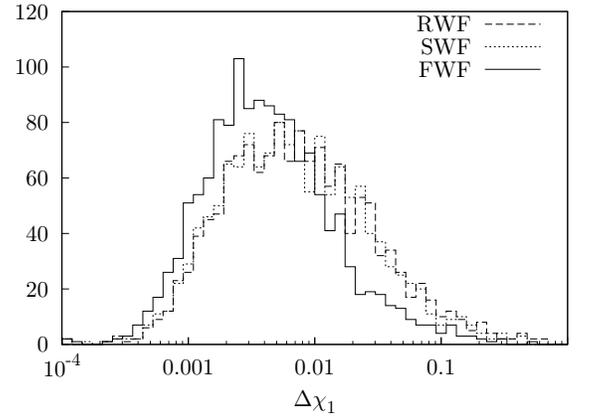}
  \caption{Estimated distribution of the measurement error on
 $\chi_1$ for a low equal-mass binary system with $m_1 = m_2 = 3 \cdot 10^5
M_\odot$. The improvement on the median value is a factor
$1.5 \upto 2$ and on the $95\%$ quantile a factor $2 \upto 4$ for the FWF
with respect to the other two waveforms.\label{fig:ch3535}}
\end{figure}

\begin{figure}[!ht]
  \includegraphics[width=\columnwidth]{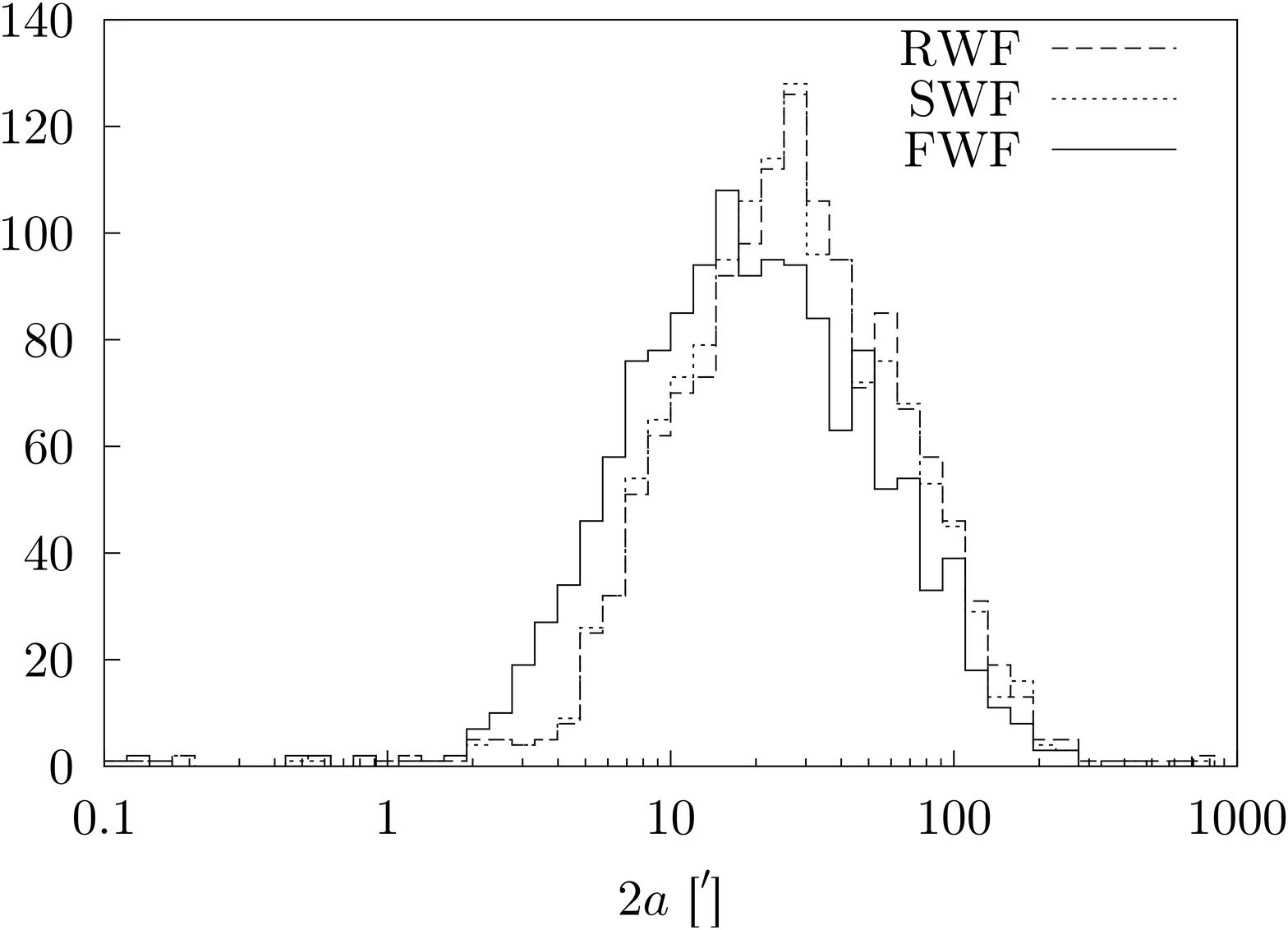}
  \caption{Estimated distribution of the major axis of the positioning error
ellipse for a low equal-mass binary system with $m_1 = m_2 = 3 \cdot 10^5
M_\odot$. We expect the error to be $\sim 1.5$ times
lower using the FWF than using the RWF.\label{fig:2a3535}}
\end{figure}

\begin{figure}[!ht]
  \includegraphics[width=\columnwidth]{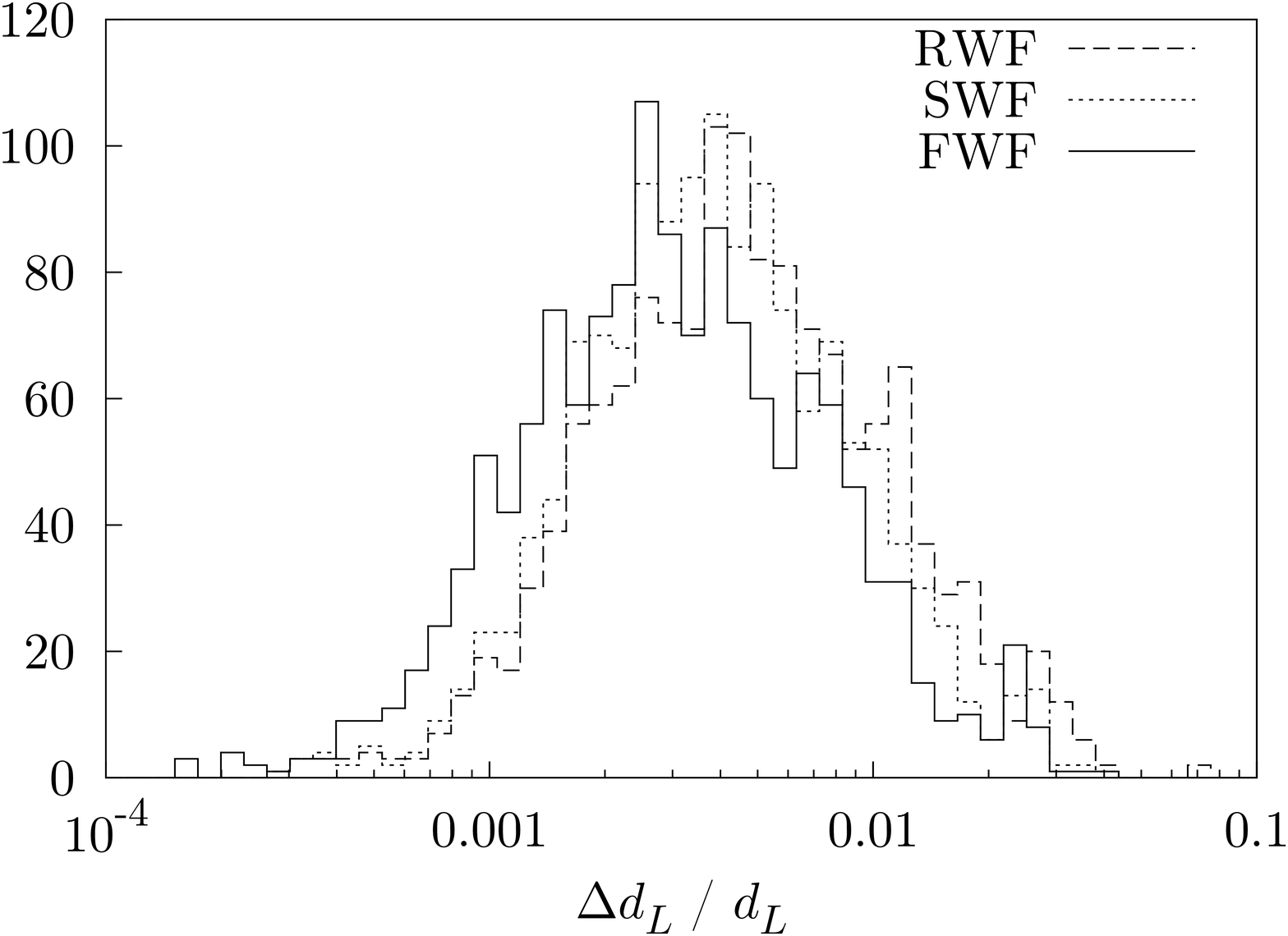}
  \caption{Estimated distribution of the measurement error on
 $d_L$ for a low equal-mass binary system with $m_1 = m_2 = 3 \cdot 10^5
M_\odot$. We expect the error to be $\sim 1.5$ times
lower using the FWF than using the RWF.\label{fig:dl3535}}
\end{figure}

We put in this class all systems of equal-mass black holes, with total mass
smaller than $10^7 M_\odot$. We chose to present as a
representative sample systems with $m_1 = m_2 = 3 \cdot 10^5
M_\odot$. We plot the estimated distribution of the errors on $m_1$ in
Fig.~\ref{fig:m3535}, on $\chi_1$ in Fig.~\ref{fig:ch3535}, on the sky
positioning in Fig.~\ref{fig:2a3535}, and on $d_L$ in
Fig.~\ref{fig:dl3535}.

In these cases, the errors on extrinsic parameters are, as for unequal-mass
systems, improved by a
factor $\sim 1.5$ for the FWF with respect to the RWF.
The errors on the spins are improved for the worst cases by a factor $2 \upto
4$, and
typically by a factor $1.5 \upto 2$ for the FWF with respect to the two other
waveforms.
However, the error on the individual
masses is improved typically by a factor $3.5 \upto 4.5$, and even by a factor
$10 \upto 20$ in the worst cases, comparing the FWF with the two others. Thus,
much
more information can be extracted from a measure of an equal-mass binary system
using the former waveform than one of the latter.

The SWF brings little improvement for intrinsic parameters in these cases,
because the odd harmonics are
absent from it, so that it has only two corrections to the RWF
instead
of five for unequal-mass systems.

\subsection{High-mass binaries}

\begin{figure}[!ht]
  \includegraphics[width=\columnwidth]{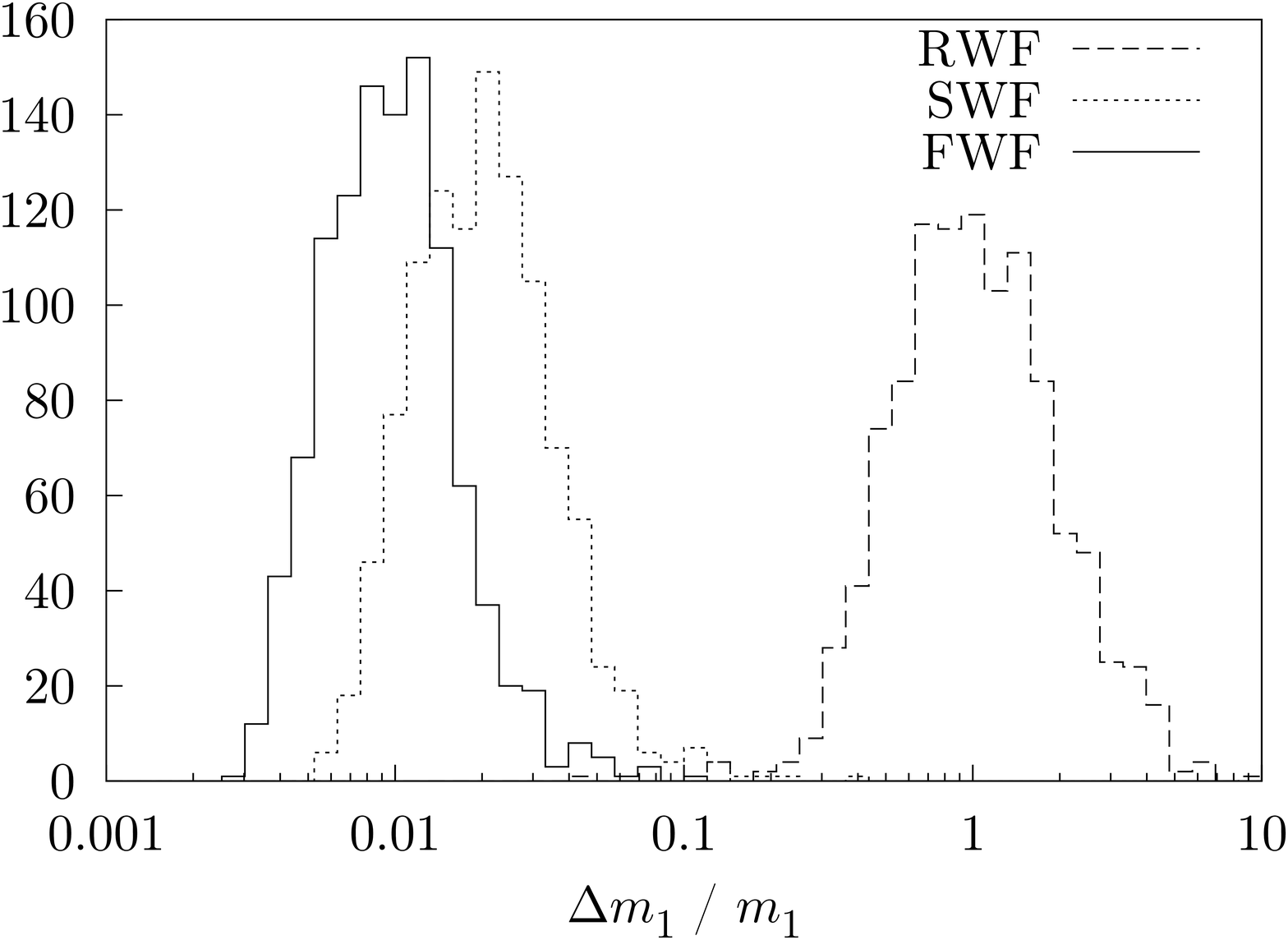}
  \caption{Estimated distribution of the measurement error on
 $m_1$ for a high-mass binary system with $m_1 = 3 \cdot 10^7
M_\odot$ and $m_2 = 10^7 M_\odot$. Very few systems are
measurable with the RWF with a precision less than $50\%$,
whereas the other two waveforms provide in the worst cases a
few percent precision, the FWF typically a factor of $1.5 \upto 2$
better than the SWF.\label{fig:m3717}}
\end{figure}

\begin{figure}[!ht]
  \includegraphics[width=\columnwidth]{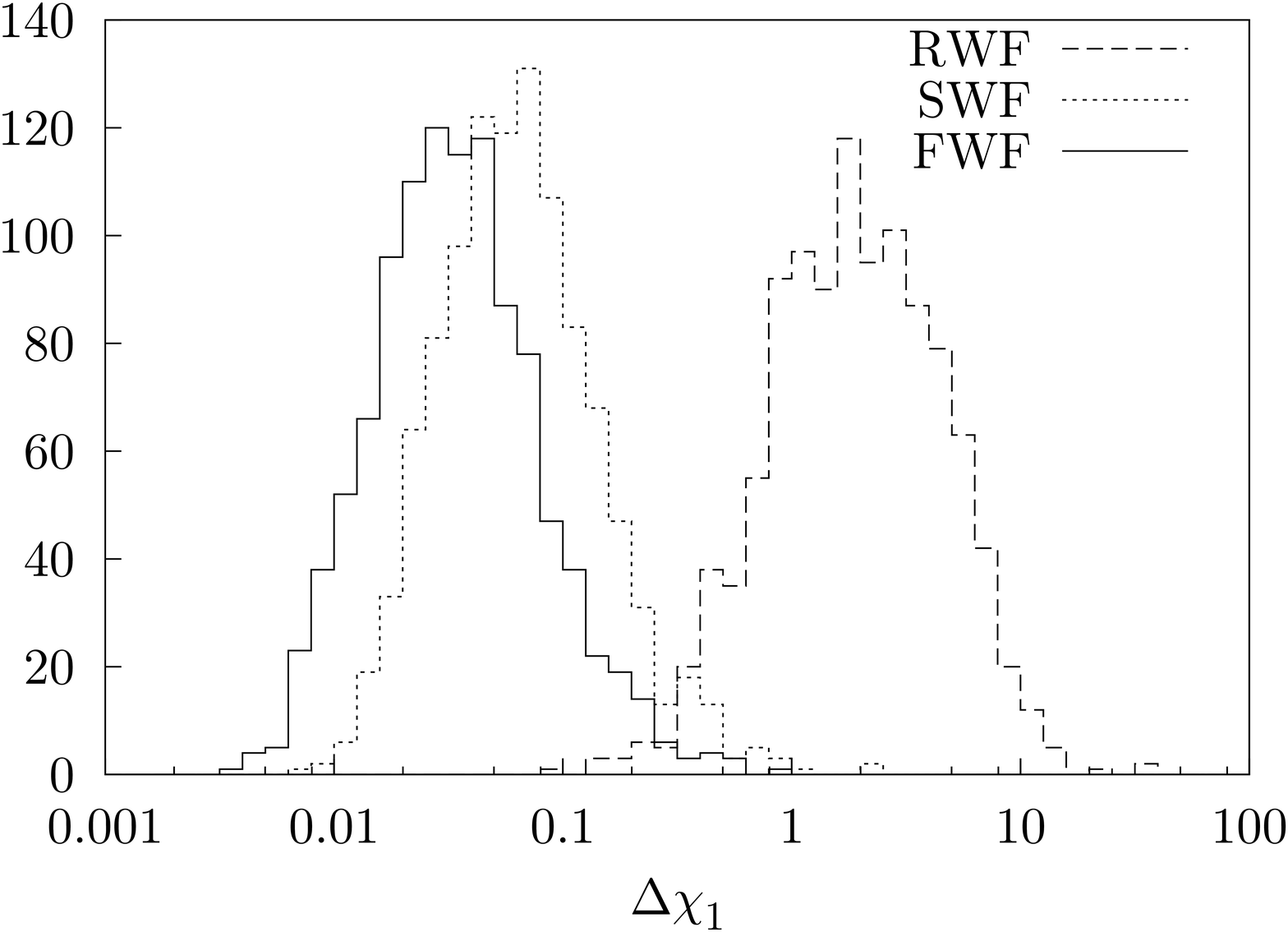}
  \caption{Estimated distribution of the measurement error on
 $\chi_1$ for a high-mass binary system with $m_1 = 3 \cdot 10^7
M_\odot$ and $m_2 = 10^7 M_\odot$. We can see that no information on the
spins can be extracted with the RWF, whereas some can be extracted with the two
others in all cases, a factor of two better for the FWF than for the
SWF.\label{fig:ch3717}}
\end{figure}

\begin{figure}[!ht]
  \includegraphics[width=\columnwidth]{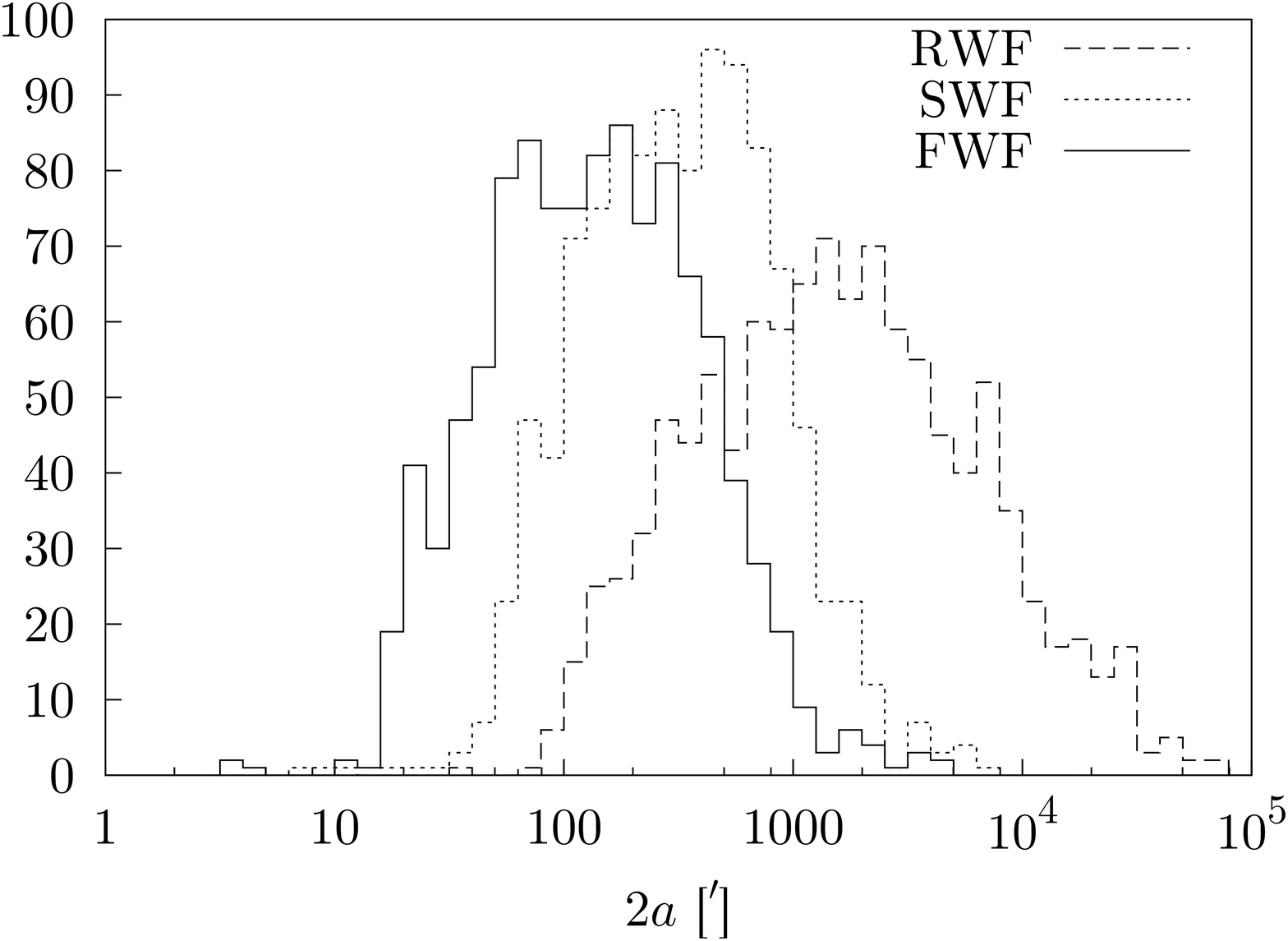}
  \caption{Estimated distribution of the major axis of the positioning error
ellipse for a high-mass binary system with $m_1 = 3 \cdot 10^7
M_\odot$ and $m_2 = 10^7 M_\odot$. We expect to have a positioning error in
the best cases ($5\%$ quantile) of $2.8^\circ$ for the
RWF, of $1^\circ$ for the SWF, and of $25'$
for the FWF.\label{fig:2a3717}}
\end{figure}

\begin{figure}[!ht]
  \includegraphics[width=\columnwidth]{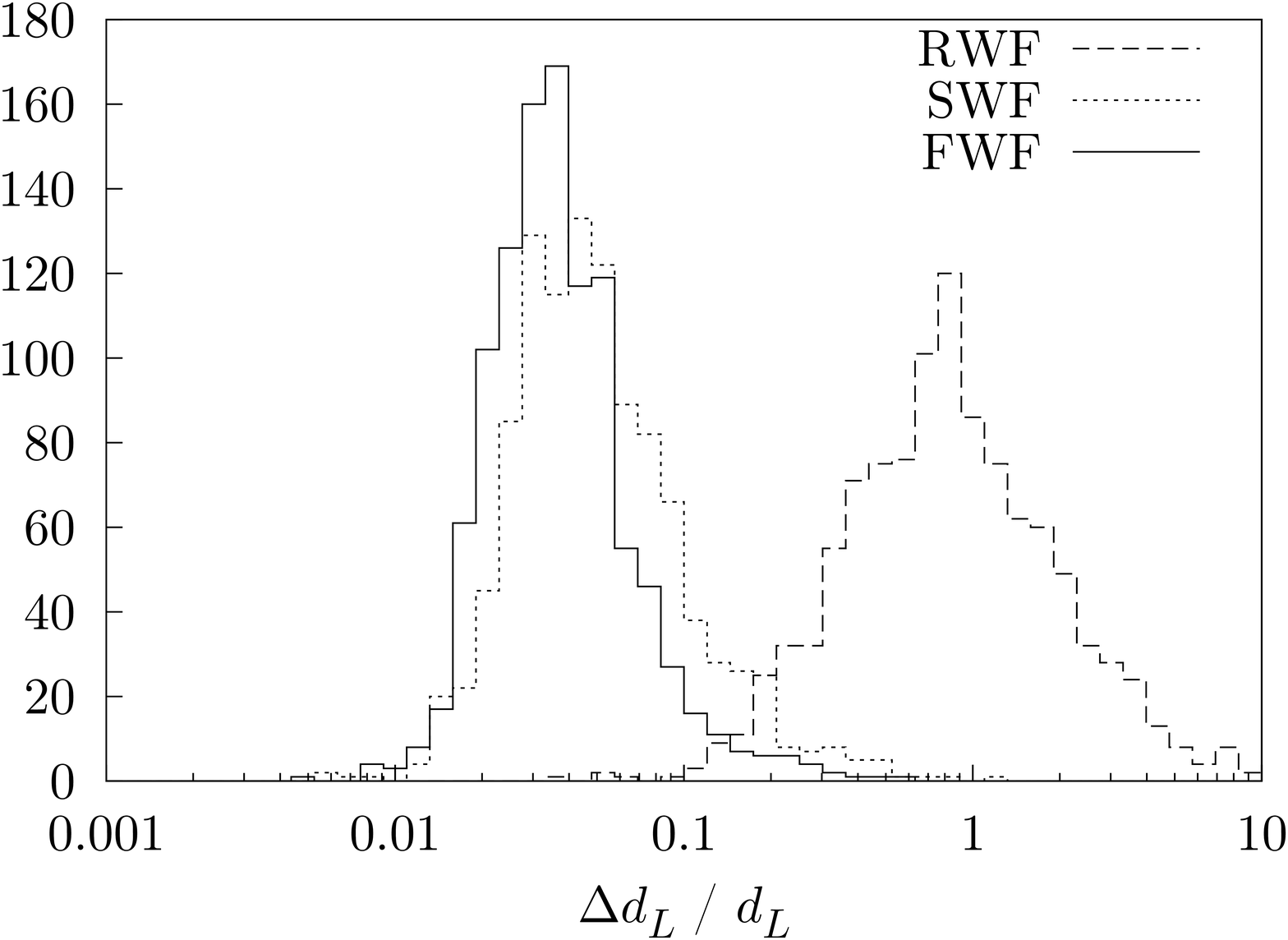}
  \caption{Estimated distribution of the measurement error on
 $d_L$ for a high-mass binary system with $m_1 = 3 \cdot 10^7
M_\odot$ and $m_2 = 10^7 M_\odot$. We do not expect a
measurement more accurate than $20\%$ to be possible with the
RWF, whereas the accuracy should be always less than $20\%$ with the SWF, and
less than $10\%$ with the FWF.\label{fig:dl3717}}
\end{figure}

We put in this class all systems with total mass higher than $10^7 M_\odot$. We
chose to present as a
representative sample systems with $m_1 = 3 \cdot 10^7 M_\odot$ and $m_2 = 10^7
M_\odot$. We plot the estimated distribution of the errors on $m_1$ in
Fig.~\ref{fig:m3717}, on $\chi_1$ in Fig.~\ref{fig:ch3717}, on the sky
positioning in Fig.~\ref{fig:2a3717}, and on $d_L$ in
Fig.~\ref{fig:dl3717}.

For this class of binaries, the second harmonic is hardly or not at all visible
in the LISA band, so that the RWF fails to provide good accuracy unless the
total mass is close to $10^7 M_\odot$. However, the SWF and FWF still provide
relatively high precision measurement on the masses, spins and luminosity
distance of a system with $m_1 = 3 \cdot 10^7 M_\odot$ at redshift one.

For equal-mass systems in this class, the FWF provides in all cases an
improvement of a factor $10 \upto 30$ for the determination of the masses with
respect to the SWF. For other parameters and/or other mass ratios, the
improvement using the FWF is of a factor $1.5 \upto 2$ with respect to the SWF.
The
fact that the SWF seems to give better results than
the FWF for the highest-mass systems comes from the fact that the SNR for
systems in this mass range is
higher with the former than with the latter. However, we do not expect to
extract
more information from a more approximate waveform.

Furthermore, some information can still be extracted from binaries that are
completely invisible to the RWF using higher harmonics.

\subsection{Upper mass limit}

We present here for all samples, the proportion of systems for
which both
individual masses can be measured at the level of $25\%$ and $50\%$ at a
redshift of $z=1$ in Table~\ref{tab:mofz}, as well as
the luminosity distance in Table~\ref{tab:dlofz}. When at
least $25\%$ accuracy is obtainable for all systems of a sample, we do not
show it on the table.

\begin{table}[!ht]
\begin{tabular}{|c|c|c|c|c|c|c|c|}
 \hline
 $m_1 [M_\odot]$ & $m_2 [M_\odot]$ & \multicolumn{3}{c|}{$25\%$} &
\multicolumn{3}{c|}{$50\%$} \\
 \hline 
 & & RWF & SWF & FWF & RWF & SWF & FWF \\
 \hline
 $10^7$ & $3 \cdot 10^6$ & $98\%$ & $100\%$ & $100\%$ & $100\%$ & $100\%$ &
$100\%$ \\
 \hline
 $10^7$ & $10^7$ & $50\%$ & $75\%$ & $100\%$ & $71\%$ & $89\%$ &
$100\%$ \\
 \hline
 $3 \cdot 10^7$ & $10^7$ & $1\%$ & $100\%$ & $100\%$ & $5\%$ & $100\%$ &
$100\%$ \\
 \hline
 $3 \cdot 10^7$ & $3 \cdot 10^7$ & $0\%$ & $0\%$ & $84\%$ & $0\%$ &
$3\%$ & $98\%$ \\
 \hline
 $10^8$ & $10^7$ & $0\%$ & $4\%$ & $2\%$ & $0\%$ & $21\%$ &
$15\%$ \\
 \hline
 $10^8$ & $3 \cdot 10^7$ & $0\%$ & $0\%$ & $0\%$ & $0\%$ & $0\%$ &
$0\%$ \\
 \hline
\end{tabular}
\caption{Proportion of the systems in all samples where both individual masses
can be determined with an accuracy better than $25\%$, resp.
$50\%$.\label{tab:mofz}}
\end{table}

\begin{table}[!ht]
\begin{tabular}{|c|c|c|c|c|c|c|c|}
 \hline
 $m_1 [M_\odot]$ & $m_2 [M_\odot]$ & \multicolumn{3}{c|}{$25\%$} &
\multicolumn{3}{c|}{$50\%$} \\
 \hline 
 & & RWF & SWF & FWF & RWF & SWF & FWF \\
 \hline
 $10^7$ & $10^7$ & $94\%$ & $99\%$ & $100\%$ & $99\%$ & $100\%$ &
$100\%$ \\
 \hline
 $3 \cdot 10^7$ & $10^7$ & $8\%$ & $97\%$ & $99\%$ & $28\%$ & $100\%$ &
$100\%$ \\
 \hline
 $3 \cdot 10^7$ & $3 \cdot 10^7$ & $0\%$ & $12\%$ & $9\%$ & $0\%$ &
$40\%$ & $47\%$ \\
 \hline
 $10^8$ & $10^7$ & $0\%$ & $5\%$ & $1\%$ & $0\%$ & $29\%$ &
$10\%$ \\
 \hline
 $10^8$ & $3 \cdot 10^7$ & $0\%$ & $0\%$ & $0\%$ & $0\%$ & $2\%$ &
$0\%$ \\
 \hline
\end{tabular}
\caption{Proportion of the systems in all samples where the luminosity distance
can be determined with an accuracy better than $25\%$, resp.
$50\%$.\label{tab:dlofz}}
\end{table}

We see in the tables that the RWF reaches its limits for $10^7 M_\odot$
binaries, whereas the SWF and FWF can still provide significant information for
$3
\cdot 10^7 M_\odot$ binaries, and even for some $10^8 M_\odot$ binaries with
high enough mass ratio.

\begin{figure}[!ht]
  \includegraphics[width=\columnwidth]{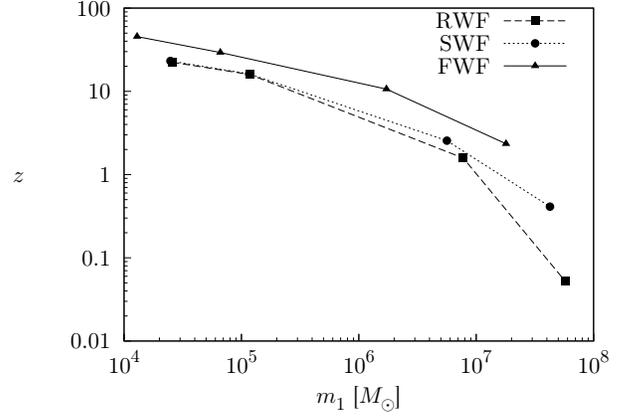}
  \caption{Maximum redshift at which a system is observable as a
function of the mass of the black holes, for equal-mass systems.
The FWF allows to observe $10^4 \upto 10^5 M_\odot$ binaries up to $z = 30
\upto 50$, the
other two up to $z = 15 \upto 25$. A binary of $\sim 2 \cdot 10^7 M_\odot$ black
holes
should be observable up to $z \approx 2$ with the FWF, $z \approx 1$ with
the SWF, and  $z \approx 0.3$ with
the RWF.\label{fig:Mq1}}
\end{figure}

\begin{figure}[!ht]
  \includegraphics[width=\columnwidth]{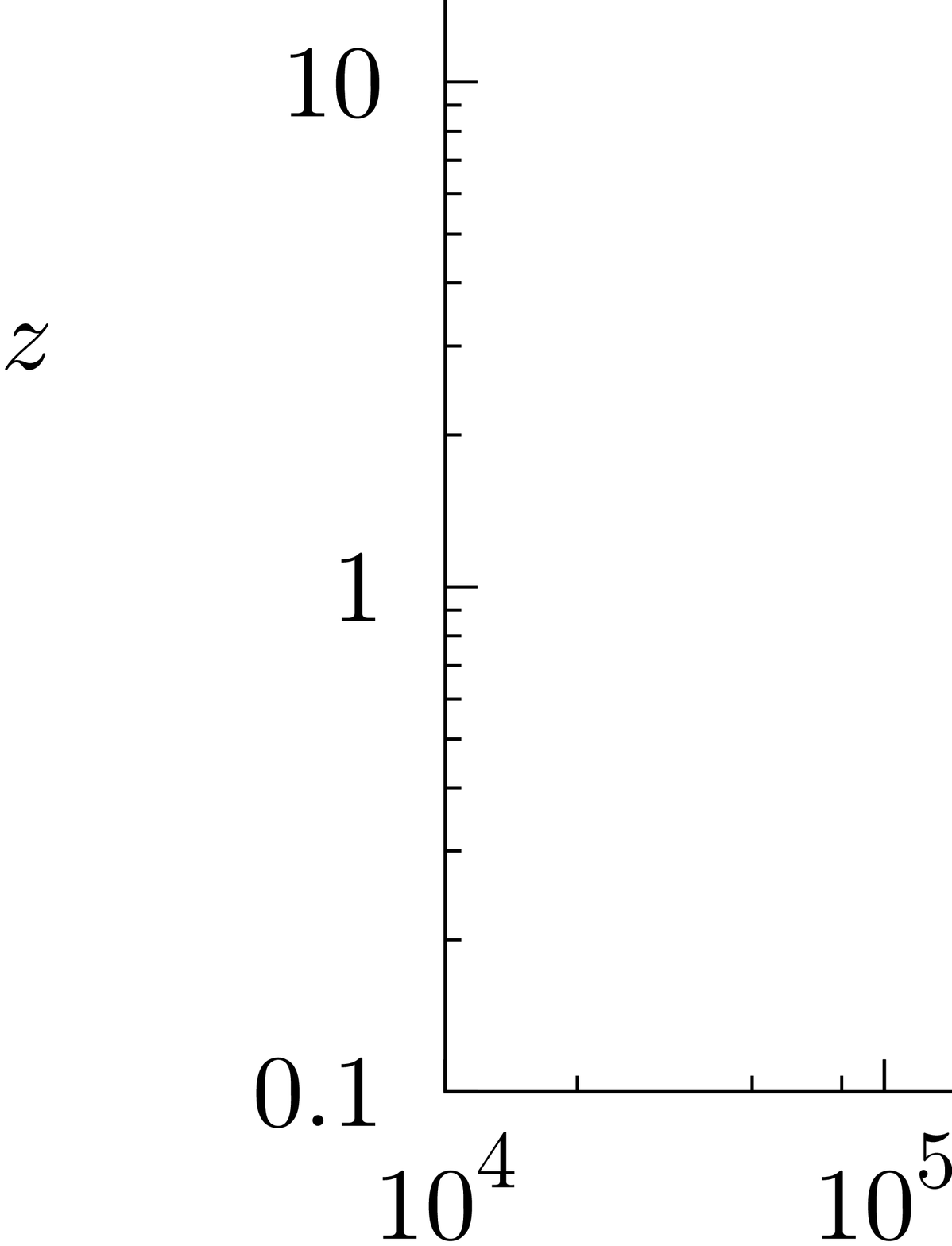}
  \caption{Maximum redshift at which a system is observable as a
function of the mass of the most massive black hole, for systems with a mass
ratio between 1:3
and 3:10. A binary with $m_1 \approx 2 \cdot 10^4 M_\odot$ should be
observable up to $z \approx 36$ with the FWF, up to $z \approx 30$ with the SWF,
up to $z \approx 28$ with the RWF. A binary with $m_1 \approx 10^7
M_\odot$ should be
observable up to $z \approx 7$ with the FWF, up to $z \approx 6$ with the SWF,
and
up to $z \approx 2$ with the RWF.\label{fig:Mq3}}
\end{figure}

\begin{figure}[!ht]
  \includegraphics[width=\columnwidth]{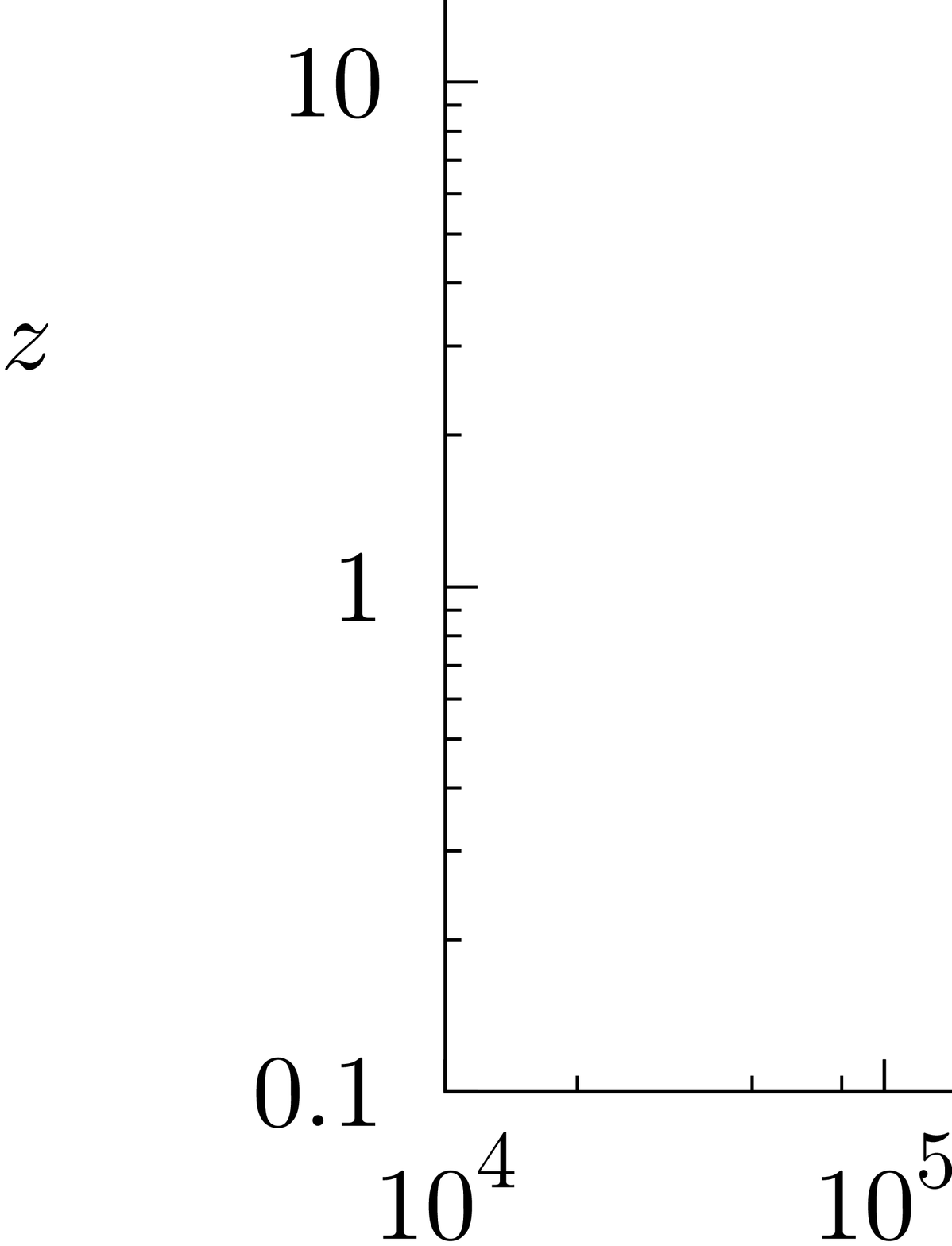}
  \caption{Maximum redshift at which a system is observable as a
function of the mass of the most massive black hole, for systems with a mass
ratio of 1:10. A binary with $m_1 \approx 6 \cdot 10^4 M_\odot$ should be
observable up to $z \approx 38$ with the FWF, up to $z \approx 34$ with the SWF,
up to $z \approx 29$ with the RWF. A binary with $m_1 \approx 10^8
M_\odot$ should still be
observable up to $z \approx 0.7$ with the FWF and SWF,
and not visible at all with the RWF.\label{fig:Mq10}}
\end{figure}

Furthermore, we computed from our simulations the maximum redshift at which a
binary system is observable, as a function of $m_1$, for different
values of the mass ratio. We chose to call a system with parameters
$(m_1,m_2,z)$
observable if at least half of the systems of these masses at this redshift have
both individual masses measurable with at least 25\% precision. We present in
Fig.~\ref{fig:Mq1} the maximum redshift at which equal-mass systems can be
observed, in Fig.~\ref{fig:Mq3} the same for
systems with a mass ratio between 1:3 and
3:10, and in Fig.~\ref{fig:Mq10} the same for systems with a mass ratio of
1:10. Some points are absent of the plots, because no signal at all could be
extracted from the RWF when the higher-mass black hole had a redshifted mass
$m_1 \approx 10^8 M_\odot$.

The figures show that a much higher redshift can be reached with the FWF than
with the other waveforms, and that the difference is bigger for mass
ratios closer to 1:1. The FWF is giving the possibility to observe any binary
system with
total mass of $M \leqslant 10^7 M_\odot$ up to a redshift of $z \gtrsim 10$,
whereas the other waveforms fail, especially for equal-mass systems. The same
combinations of redshifted masses can be observed with the FWF at redshifts
$1.5 \upto 5$ times higher than with the RWF, which could greatly help
constraining
black hole and galaxy formation models~\cite{svh}.

\subsection{Extrinsic parameters}

\begin{figure}[!ht]
  \includegraphics[width=\columnwidth]{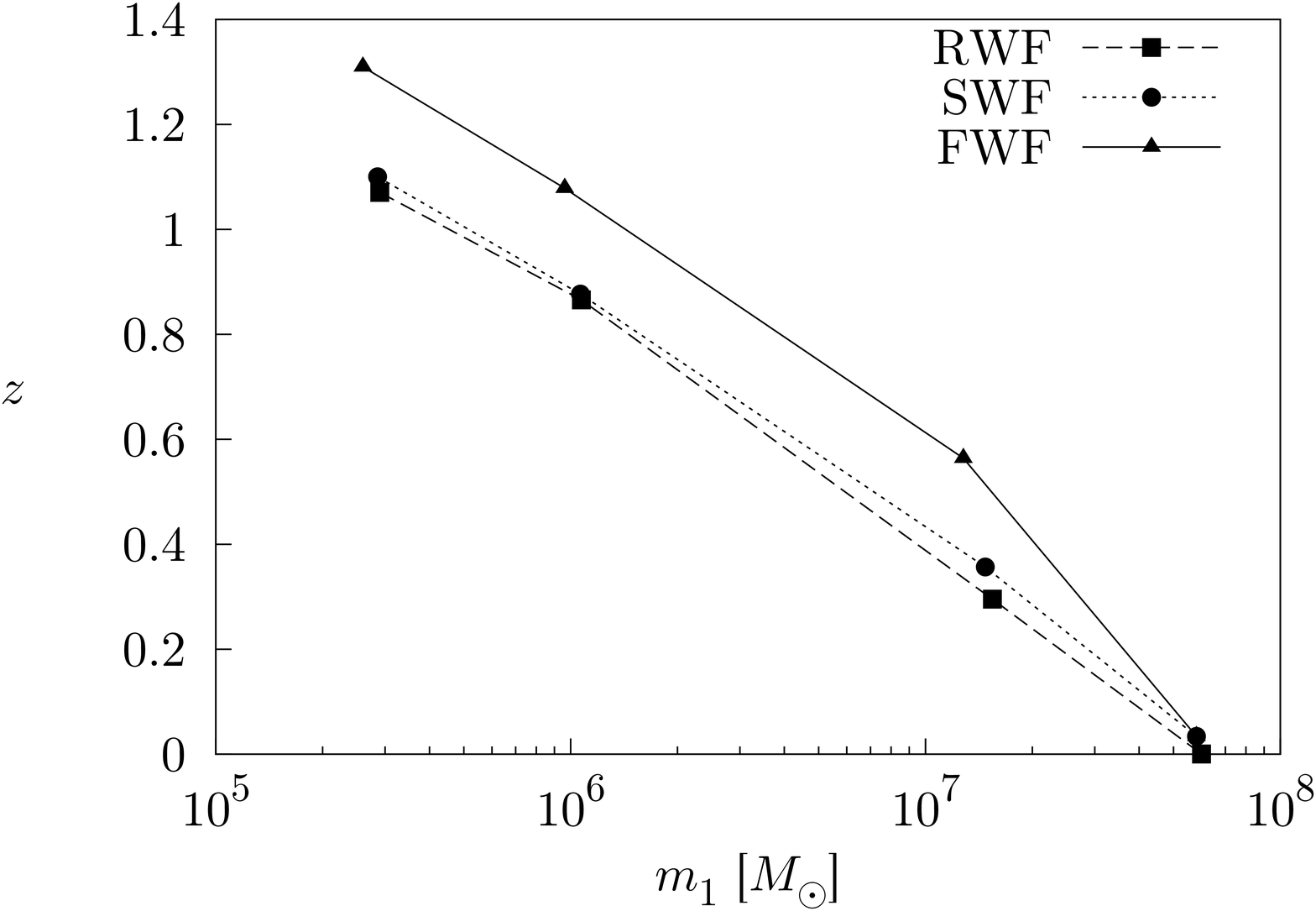}
  \caption{Maximum redshift at which the binary can be located with a $30'$
precision as
function of the mass of the most massive black hole, for equal-mass systems.
The FWF allows to locate a binary accurately up to a redshift
of $\sim 0.2$ greater than the two other waveforms.\label{fig:aq1}}
\end{figure}

\begin{figure}[!ht]
  \includegraphics[width=\columnwidth]{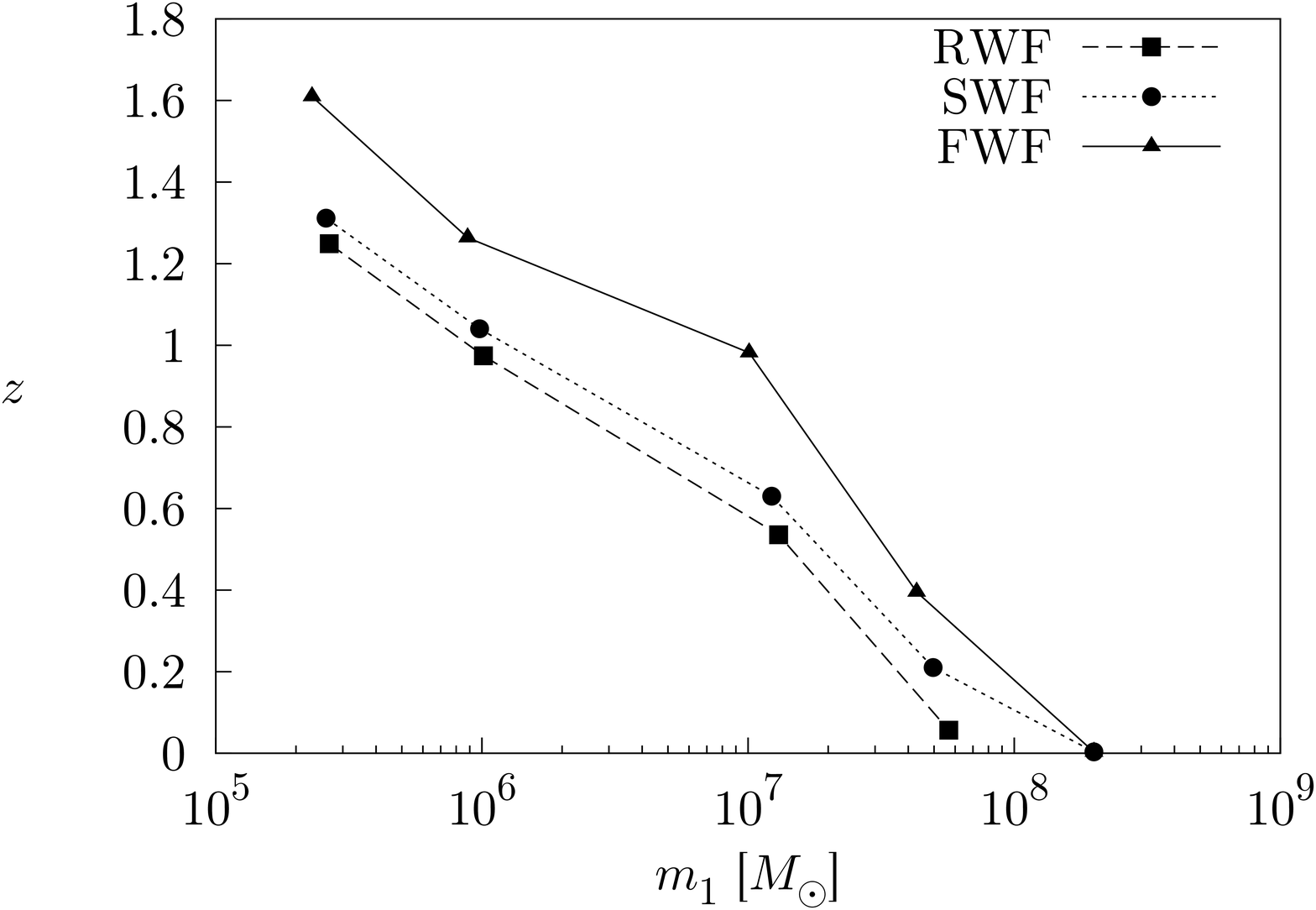}
  \caption{Maximum redshift at which the binary can be located with a $30'$
precision as a
function of the mass of the most massive black hole, for systems with a mass
ratio between 1:3 and 3:10. The FWF allows to locate a binary accurately up to a
redshift
of $0.2 \upto 0.3$ greater than the SWF, and the SWF up to a redshift
of $\sim 0.1$ greater than the RWF.\label{fig:aq3}}
\end{figure}

\begin{figure}[!ht]
  \includegraphics[width=\columnwidth]{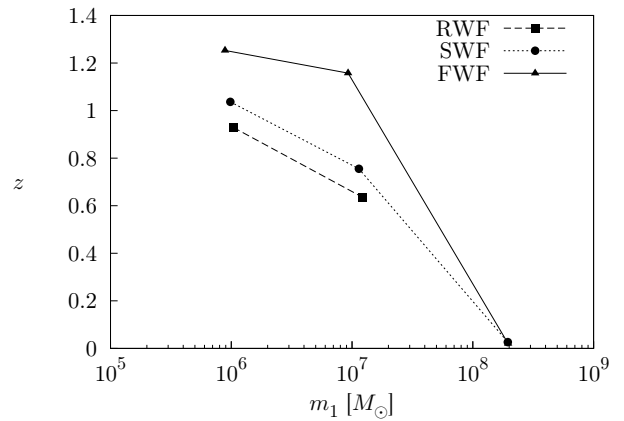}
  \caption{Maximum redshift at which the binary can be located with a $30'$
precision as a
function of the mass of the most massive black hole, for systems with a mass
ratio of 1:10.  The FWF allows to locate a binary accurately up to a
redshift
of $0.2 \upto 0.4$ greater than the SWF, and the SWF up to a redshift
of $\sim 0.1$ greater than the RWF.\label{fig:aq10}}
\end{figure}

We plot here as a function of the mass of the most massive black hole the
maximum redshift where the major axis of the positioning error ellipse is less
than
$30'$ for half of the binaries. Equal-mass binaries are shown in
Fig.~\ref{fig:aq1}, binaries with mass ratio between 1:3 and 3:10 in
Fig.~\ref{fig:aq3}, and binaries with a mass ratio of 1:10 in
Fig.~\ref{fig:aq10}.

We found that in all cases, the localization in the sky is far more difficult
than
the determination of the luminosity distance. Irrespective of the waveforms,
masses and mass ratios, the luminosity distance can be measured with a
precision of $0.3\% \upto 0.5\%$ when the major axis of the positioning error
ellipse is
$30'$.

The FWF could help locating binaries accurately
enough for the observation of their merger to become possible up to redshifts
$0.2 \upto 0.4$ greater than the two other waveforms. The SWF could furthermore,
in
the case of unequal-mass binaries, go up to redshifts $\sim 0.1$ greater than
the RWF.

Our simulations show that supermassive black hole binaries could be very
accurate standard candles, and could successfully extend the measurements of
the Hubble diagram up to redshifts of $z=1.6$, with a precision on the
luminosity distance of a few per mille. This would be a great breakthrough in
the distance ladder, as the current most effective standard candles at large
distances, type Ia supernovae, provide much less precision on large luminosity
distances.

\section{Conclusion}

\label{sec:conclusion}

The fact that the detection of gravitational waves with interferometric
detectors relies on template-based searches should suggest to use the most
accurate waveform available for detecting systems emitting such waves.
The gravitational waveform of two spinning bodies orbiting each
other is however complicated, and each new further step implies more
and more complicated corrections to the waveform. It is a good thing to know
whether it is worth using a more accurate waveform, and in what cases.

Comparing our results for the RWF to those of Lang and Hughes~\cite{langhughes},
we found
that our estimates are systematically a factor of $\sim 1.2$ more pessimistic
for lower-mass binaries, and a factor of $\sim 2$ more pessimistic for
higher-mass binaries. As stated in section~\ref{sec:dataal}, we expected such a
discrepancy because of the differences in the noise curves we used. As
higher-mass binaries spend more time in the lower frequency
range where the noise differs the most, we expected the differences for these
binaries to be larger than for less massive binaries.

We found that the addition of higher harmonics to the waveform at the 2PN level
can help increasing the mass limit above which no information can be extracted
from
the signal. The amplitude corrections can also bring important improvements to
the determination of the individual masses, also for lower-mass binaries.
The FWF allows for detecting binaries up to redshifts $z > 40$, whereas the
other waveforms can allow detection up to $z \approx 30$. This could be very
interesting for constraining galaxy and black hole formation models.

The range of LISA could be also extended for the determination of the Hubble
diagram. The RWF would allow to measure the redshift-luminosity distance
relation at a few per mille precision up to $z \approx 1.2$, whereas the FWF
would allow measures with the same precision up to $z \approx 1.6$. It would be
interesting to quantify how well LISA would be able to determine the Hubble
diagram.

The use of the full waveform as a template for the gravitational
radiation of comparable-mass binaries can be important to extract the maximum
information possible, especially for high-mass and/or close to equal-mass
binaries. However, in the case of unequal low-mass binaries at low redshifts,
the restricted waveform used in earlier studies can be sufficient.
The fact that using the full waveform in these searches can fail to provide
much more accuracy for some systems suggests that including more parameters,
such as eccentricity~\cite{yabw} or alternative gravity
parameters~\cite{bbw,arunwill,stavridiswill}, could keep the accuracy for the
other
parameters reasonably high, allowing to extract more information from the
detection of a wave. Gravitational waves can be a powerful tool for
constraining alternative theories of gravitation, in the sense that each event
will provide an independent measurement of their parameters.

Even though the spin-coupling effects in the wave amplitude are not yet known at
the 2.5PN level, it could be interesting to compare the measurement accuracy we
get from a 2.5PN accurate waveform as compared to the 2PN accurate one we used
in this study. It has been shown~\cite{cutlervallisneri,umstaettertinto} in the
case of spinless bodies that theoretical errors due to the inaccuracy of the
waveform can be important for high SNR systems. It would also be interesting to
perform the same study for spinning systems.

\begin{acknowledgments}
We would like to thank Neil Cornish for useful precisions about noise models
for interferometric gravitational wave detectors, and Prasenjit Saha for
useful discussions about the efficiency and precision of numerical methods. We
would also like to thank the referee for useful comments. A.~K.
and M.~S. are supported by the Swiss National Science Foundation.
\end{acknowledgments}

\appendix*

\section{Polarizations}

We give here the plus and cross polarizations we used in our studies, in terms
of the orbital phase $\psi$. The plus and cross polarizations of the simplified
waveform (SWF) is obtained by keeping only the lowest order in
$A_{+}^{(n)}$ and $B_{\times}^{(n)}$, and those of the RWF by keeping
only the lowest order of $A_{+}^{(2)}$ and $B_{\times}^{(2)}$. To
obtain the SWF and RWF, the function $S(f)$ in Eq.~\eqref{htilde} should
also be set
to $S(f)=1$.

The plus and cross polarization waveforms are:
\begin{align}
 h_{+,\times} &= \frac{2GM\nu x}{d_L c^2} \left[ \sum_{n \geqslant 0}
\left(A_{+,\times}^{(n)} \cos n\psi + B_{+,\times}^{(n)} \sin n\psi \right)
\right], \\
 s_i &= \norm{\uvec{L} \times \uvec{n}}, \\
 c_i &= \uvec{L} \cdot \uvec{n}.
\end{align}

With the use of the spin-orbit coupling parameter $\tau$ defined as:
\begin{equation}
 \tau \equiv \frac{c}{G M} \left( \frac{\bm{S}_1}{m_1} - \frac{\bm{S}_2}{m_2}
\right) \cdot \uvec{L},
\end{equation} 
we can write the nonvanishing parameters $A_{+,\times}^{(n)}$ and
$B_{+,\times}^{(n)}$, as currently known at 2PN level. The value of
$\bar{\omega}$
appearing below can be chosen arbitrarily.

\begin{widetext}

\begin{subequations}
\begin{align}
 A_+^{(0)} &= - \frac{s_i^2}{96} \left( 17 + c_i^2 \right)\\
 A_+^{(1)} &= s_i \sqrt{1 - 4\nu} \left( -\frac{5}{8} - \frac{c_i^2}{8} \right)
x^{1/2} + s_i \tau x \nonumber\\
&\qquad\qquad + s_i \sqrt{1 - 4\nu} \left[ \frac{19}{64} +
\frac{5c_i^2}{16} - \frac{c_i^4}{192} + \nu \left( -\frac{49}{96} +
\frac{c_i^2}{8}
+
\frac{c_i^4}{96} \right) \right] x^{3/2}  + s_i \sqrt{1 - 4\nu} \left(
-\frac{5}{8} - \frac{c_i^2}{8}
\right) \pi x^2 \\
 A_+^{(2)} &= \left( - 1 - c_i^2 \right) + \left[ \frac{19}{6}
+ \frac{3c_i^2}{2} - \frac{c_i^4}{3} + \nu \left( - \frac{19}{6} +
\frac{11c_i^2}{6} + c_i^4 \right) \right] x + \left[ 2\pi \left( - 1 - c_i^2
\right) + \frac{8}{3}
\beta\left( 3 - 9 c_i^2, 2 - 10 c_i^2 \right) \right] x^{3/2} \nonumber\\
 &\qquad\qquad\qquad + \bigg[ \frac{11}{60} + \frac{33c_i^2}{10} +
\frac{29c_i^4}{24} -
\frac{c_i^6}{24} + \nu \left( \frac{353}{36} - 3 c_i^2 - \frac{251c_i^4}{72} +
\frac{5c_i^6}{24} \right) \nonumber\\
&\qquad\qquad\qquad\qquad\qquad\qquad\qquad + \nu^2 \left( -\frac{49}{12} +
\frac{9c_i^2}{2} -
\frac{7c_i^4}{24} - \frac{5c_i^6}{24} \right) - 2 \sigma\left( 1 + c_i^2, 0
\right) \bigg] x^2 \\
 A_+^{(3)} &= s_i \sqrt{1 - 4 \nu} \left( \frac{9}{8} +
\frac{9c_i^2}{8} 
\right) x^{1/2} \nonumber\\
 &\quad + s_i \sqrt{1 - 4\nu} \left[- \frac{657}{128} -
\frac{45c_i^2}{16} + \frac{81c_i^4}{128} + \nu \left( \frac{225}{64} -
\frac{9c_i^2}{8} - \frac{81c_i^4}{64} \right) \right] x^{3/2} + s_i \sqrt{1 -
4\nu} \left( \frac{27}{8} + \frac{27c_i^2}{8}
\right) \pi x^2 \\
 A_+^{(4)} &= s_i^2\left( 1 + c_i^2 \right) \left( - \frac{4}{3} +4\nu \right)
x \nonumber\\
 &\quad + \left[ \frac{118}{15} - \frac{16c_i^2}{5} - \frac{86c_i^4}{15} +
\frac{16c_i^6}{15} + \nu \left( - \frac{262}{9} + 16 c_i^2 + \frac{166c_i^4}{9}
-
\frac{16c_i^6}{3} \right) + \nu^2 \left( 14 - 16c_i^2 - \frac{10c_i^4}{3} +
\frac{16c_i^6}{3} \right) \right] x^2\\
 A_+^{(5)} &= s_i^3 \sqrt{1 - 4 \nu} \left( \frac{625}{384} -
\frac{625\nu}{192} \right) \left( 1 + c_i^2 \right) x^{3/2} \\
 A_+^{(6)} &= s_i^4 \left( 1 + c_i^2 \right) \left( -\frac{81}{40} +
\frac{81\nu}{8} - \frac{81\nu^2}{8} \right) x^2
\end{align}
\end{subequations}

\begin{subequations}
\begin{align}
 A_\times^{(1)} &= s_i c_i \sqrt{1 - 4\nu} \left[ -\frac{9}{20} -
\frac{3\log2}{2} + \frac{9}{4} \log\left( \frac{\omega}{\bar{\omega}} \right)
\right] x^2 \\
 A_\times^{(2)} &= 12c_i \log\left( \frac{\omega}{\bar{\omega}} \right) x^{3/2}
\\
 A_\times^{(3)} &= s_i c_i \sqrt{1 - 4\nu} \left[ \frac{189}{20} -
\frac{27\log(3/2)}{2} - \frac{81}{4} \log\left( \frac{\omega}{\bar{\omega}}
\right)
\right] x^2
\end{align}
\end{subequations}

\begin{subequations}
\begin{align}
 B_\times^{(1)} &= -\frac{3}{4} s_i c_i \sqrt{1 - 4\nu} \, x^{1/2} + s_i c_i
\tau x + s_i c_i \sqrt{1 - 4\nu} \left[ \frac{21}{32} - \frac{5c_i^2}{96}
+ \nu\left( -\frac{23}{48} + \frac{5c_i^2}{48} \right) \right]
x^{3/2} - \frac{3\pi}{4} s_i c_i \sqrt{1 - 4\nu} \, x^2 \\
 B_\times^{(2)} &= -2 c_i + c_i \left[ \frac{17}{3} - \frac{4c_i^2}{3} +
\nu\left( -\frac{13}{3} + 4 c_i^2 \right) \right] x + c_i \left[ -4\pi -
\frac{4}{3} \beta\left( 1 + 3 c_i^2, 3
c_i^2 \right) \right] x^{3/2} \nonumber\\
 &\qquad\qquad + c_i \left[ \frac{17}{15} + \frac{113c_i^2}{30} -
\frac{c_i^4}{4} + \nu \left( \frac{143}{9} - \frac{245c_i^2}{18}
+\frac{5c_i^4}{4} \right) + \nu^2 \left( -\frac{14}{3} + \frac{35c_i^2}{6} -
\frac{5c_i^4}{4} \right) - 4 \sigma\left( 1, 0 \right) \right] x^2 \\
 B_\times^{(3)} &= \frac{9}{4} s_i c_i \sqrt{1 - 4\nu} \, x^{1/2} + s_i c_i
\sqrt{1 - 4\nu} \left[ - \frac{603}{64} + \frac{135c_i^2}{64} + \nu \left(
\frac{171}{32} - \frac{135c_i^2}{32} \right) \right]
x^{3/2} + \frac{27\pi}{4} s_i c_i \sqrt{1 - 4\nu} \, x^2 \\
 B_\times^{(4)} &= c_i s_i^2 \left( -\frac{8}{3} + 8\nu \right) x + c_i \left[
\frac{44}{3} - \frac{268c_i^2}{15} + \frac{16c_i^4}{5} + \nu \left(
-\frac{476}{9} + \frac{620c_i^2}{9} - 16c_i^4
\right) + \nu^2 \left( \frac{68}{3} - \frac{116c_i^2}{3} + 16c_i^4 \right)
\right] x^2 \\
 B_\times^{(5)} &=  s_i^3 c_i \sqrt{1 - 4\nu} \left( \frac{625}{192} -
\frac{625\nu}{96} \right) x^{3/2} \\
 B_\times^{(6)} &= s_i^4 c_i \left( -\frac{81}{20} + \frac{81\nu}{4} -
\frac{81\nu^2}{4} \right) x^2
\end{align}
\end{subequations}

\begin{subequations}
\begin{align}
 B_+^{(1)} &= s_i \sqrt{1 - 4\nu} \left[ \frac{11}{40} + \frac{5 \log2}{4} +
\left( \frac{7}{40} + \frac{\log2}{4} \right) c_i^2 + \left( -\frac{15}{8} -
\frac{3c_i^2}{8} \right)
\log\left(\frac{\omega}{\bar{\omega}}\right) \right] x^2 \\
 B_+^{(2)} &= \left( - 6 - 6 c_i^2 \right) \log\left(
\frac{\omega}{\bar{\omega}}
\right) x^{3/2} \\
 B_+^{(3)} &= s_i \sqrt{1 - 4\nu} \left[ -\frac{189}{40} +
\frac{27\log(3/2)}{4} + \frac{81}{8} \log\left( \frac{\omega}{\bar{\omega}}
\right)
\right] \left( 1 + c_i^2 \right) x^2
\end{align}
\end{subequations}

\end{widetext}

\end{document}